\documentclass[aps,prd,reprint,amsmath,amssymb,longbibliography,nofootinbib,floatfix]{revtex4-2}
\usepackage[T1]{fontenc}
\usepackage[utf8]{inputenc}
\usepackage{bm,graphicx,booktabs}
\usepackage[colorlinks=true,linkcolor=blue,citecolor=blue,urlcolor=blue,hypertexnames=false]{hyperref}
\hypersetup{pdfauthor={Xiaozhu Yu, Xinyang Wang},pdftitle={Kernel moments and thermodynamic stability in nonlocal Nambu--Jona-Lasinio models}}
\graphicspath{{figures/}}
\newcommand{\dd}{\mathrm{d}}
\newcommand{\ST}{\mathcal{S}_T}
\newcommand{\ip}{\int_{\bm p}}
\newcommand{\iq}{\int_q}
\newcommand{\Om}{\overline\Omega}
\newcommand{\MeV}{\,\mathrm{MeV}}
\newcommand{\GeV}{\,\mathrm{GeV}}
\newcommand{\sfree}{s_{\mathrm{free}}}
\newcommand{\sint}{s_{\mathrm{int}}}
\newcommand{\cfix}{s_{T}^{\mathrm{fix}}}
\begin{document}
\raggedbottom
\title{Kernel moments and thermodynamic stability in nonlocal Nambu--Jona-Lasinio models}
\author{Xiaozhu Yu}
\affiliation{Department of Physics, Jiangsu University, Zhenjiang 212013, China}
\author{Xinyang Wang}
\email[Corresponding author: ]{wangxy@aust.edu.cn}
\affiliation{Center for Fundamental Physics, School of Mechanics and Optoelectronic Physics,
Anhui University of Science and Technology, Huainan, Anhui 232001, China}
\date{October 7, 2026}
\begin{abstract}
We study thermodynamic consistency and stability in a covariant two-flavor nonlocal Nambu--Jona-Lasinio model at zero chemical potential in the homogeneous quark mean-field approximation. A controlled large-source expansion gives a sufficient entropy-sign criterion directly from Euclidean kernel moments, without locating propagator poles. At fixed couplings and kernel, every admissible positive Gaussian mixture develops negative total entropy at sufficiently large source mass for each fixed positive temperature. In a separate light-source fit to pion observables and a subtracted condensate, one quartic-exponent kernel has positive entropy and heat capacity for all $T>0$, while $dP/d\epsilon=s/C_V$ exceeds unity on a finite interval. This constructive example separates thermal positivity from the static EOS slope test. Numerical refits show that the enhancement weakens with the condensate input, and a certified counterexample excludes uniform thermal positivity over the adopted matching domain. The results provide diagnostics of the mean-field functional, rather than a preferred phenomenological kernel. Omitted thermal mesons can change the low-temperature slope test, and the static derivative does not determine a real-time front velocity.
\end{abstract}
\maketitle

\section{Introduction}\label{sec:intro}
The Nambu--Jona-Lasinio (NJL) model~\cite{Nambu1961a,Nambu1961b} provides a useful description of chiral symmetry breaking, but its thermal predictions depend on how the interaction is regulated. At zero chemical potential, Ref.~\cite{Xue2022} already demonstrated incorrect high-temperature heat-capacity behavior and superluminal sound speeds from thermal regularization, comparing hard-cutoff, soft-cutoff, and Pauli--Villars schemes. Removing thermal regularization restores the free-gas behavior in those examples, but can drive the constituent mass below the current mass. Ref.~\cite{Pasqualotto2023} subsequently examined these cutoff effects at finite quark and chiral chemical potentials, including isentropic trajectories and trace-anomaly diagnostics. Renormalization-group (RG) consistency provides a framework for controlling cutoff artifacts in low-energy effective theories~\cite{Braun2019}. Its application to local NJL and PNJL models with separate vacuum and thermal cutoff scales improves high-temperature thermodynamics, while retaining a regime with a constituent mass below the current mass and a nonmonotonic thermal-cutoff dependence of the RGPNJL sound-speed diagnostic~\cite{Tang:2026omz}. The mass crossing, also studied as superrestoration, is itself regularization dependent in the local NJL model~\cite{Inagaki2024}; it is distinct from the entropy and heat-capacity sign tests considered here. Does a finite nonlocal interaction, with unrestricted thermal integration, avoid these problems?

Nonlocal NJL models replace contact vertices by momentum-dependent form factors. Their vacuum pion sector and axial currents~\cite{BowlerBirse1995,GomezDumm2006,Noguera2008}, finite-temperature formulations, and Polyakov extensions~\cite{GomezDumm2002,GomezDumm2005,Hell2009,Contrera2010} are well established. Ultraviolet convergence does not, however, ensure thermal stability. Complex-conjugate propagator poles can generate oscillating thermal contributions~\cite{Benic2012}. The dependence of entropy oscillations on the regulator, wave-function renormalization, and the Polyakov loop was studied in Ref.~\cite{Benic2014}. A real-time analysis isolated highly unstable quasiparticle poles and examined their exclusion or avoidance as remedies in specific parameter sets~\cite{Marquez2014}. Residual instabilities also occur with subtracted propagator poles and complex-plane cuts in a modified mass function~\cite{Radzhabov2025}, although that study does not establish negative entropy. Related tests of single-valuedness and the slope of $P(\epsilon)$ have been applied to propagator-based equations of state~\cite{Canfora2016}.

We distinguish three requirements. First, condensates and entropy must follow from one source-dependent thermodynamic potential; the gap equation alone leaves a field-independent background undetermined~\cite{Moreira2012}. Second, the equilibrium thermal responses must have acceptable signs, including the response of the mean field itself. Third, even positive entropy and heat capacity need not ensure $dP/d\epsilon\leq1$. Stability in the retained scalar direction does not replace either thermal test.

We address these requirements through two separate calculations. At fixed couplings and kernel, a controlled large-current-mass expansion gives a sufficient entropy-sign criterion directly from Euclidean kernel moments. It applies to a class of positive Gaussian mixtures without locating their propagator poles. The added result is a criterion on the kernel with a controlled asymptotic remainder; a general equivalence to the established complex-pole oscillation mechanism is not proved here. The pole representation used for the nominal quartic model in Appendix~\ref{app:phase27-low-temperature} is a separate, model-specific result. The large-source limit tests the prescribed functional and is not a heavy-quark extrapolation of a vacuum fit. We then refit light-source models to pion observables and a subtracted condensate. One quartic-exponent fit constructively separates positive thermal responses for every $T>0$ from the static EOS slope test. Ordinary folds and refitted input scans describe the calibration dependence, while an exactly scaled counterexample excludes uniform positivity over the adopted matching-input domain.

Sections~\ref{sec:model}--\ref{sec:covariant} develop the common-potential and kernel-moment tests. Sections~\ref{sec:fit}--\ref{sec:inputresponse} examine the fitted light-source models, their EOS, and input sensitivity. Technical enclosures and proofs are collected in the appendices. Throughout, the thermal potential contains only the quark determinant and homogeneous scalar mean field; vacuum pion matching does not add a thermal pion gas.

\section{Sources and thermal responses from a common potential}\label{sec:model}
\subsection{Model and subtraction convention}
We use natural units and consider the model at zero chemical potential with $N_c=3$ colors, $N_f=2$ degenerate quark flavors, and $d=N_cN_f=6$. The common current-quark mass $m=m_u=m_d$ is treated as a scalar source. We retain a real, homogeneous scalar mean field $\sigma$ and set the wave-function renormalization to unity. We express all potentials as densities and use $C_V$ for the heat capacity per unit volume. At temperature $T$, we use the spatial momentum magnitude $p=|\bm p|$ and fermionic Matsubara frequencies $\omega_n=(2n+1)\pi T$, with integer $n$. The Euclidean momentum, quark mass function, and integration measure are defined by
\begin{align}
 Q^2&=p^2+\omega_n^2, & M(Q^2)&=m+\sigma g(Q^2),\\
 \ST f&=T\sum_{n\in\mathbb Z}\ip f,&
 \ip&=\int\frac{\dd^3p}{(2\pi)^3}.
\end{align}
The dimensionless form factor $g(Q^2)$, referred to as the kernel, specifies the momentum dependence of the interaction contribution $\sigma g(Q^2)$ to $M(Q^2)$. Thus $m$ is the current mass, whereas $M(Q^2)$ is the momentum-dependent quark mass function. For an even Matsubara integrand, the measure becomes $\ST f=T\pi^{-2}\sum_{n\geq0}\int_0^\infty p^2f\,\dd p$. In the examples below, $g(0)=1$, so $M(0)=m+\sigma$; $\Lambda$ is the characteristic momentum scale of the kernel.

Following the standard subtraction and free-gas restoration used in nonlocal models~\cite{GomezDumm2005}, we separate the interaction contribution as
\begin{equation}
 V(\sigma;m,T)=\frac{\sigma^2}{2G}
 -2d\ST\log\frac{Q^2+M^2(Q^2)}{Q^2+m^2}.
 \label{eq:V}
\end{equation}
Here $G>0$ is the scalar--pseudoscalar four-quark coupling, with mass dimension $-2$. We then restore the free thermal contribution at the current mass,
\begin{equation}
 B_m(T)=-4dT\ip\log(1+e^{-E_m/T}),
 \label{eq:Bm}
\end{equation}
where $E_m=\sqrt{p^2+m^2}$ is the free-quark energy. The full thermodynamic potential density before minimization is $\Omega=V+B_m+K(m)$. Minimizing over the retained real homogeneous scalar field gives $\sigma_*(m,T)$ and the equilibrium potential density
\begin{equation}
 \Om(m,T)=\Omega(\sigma_*(m,T);m,T).
\end{equation}
We similarly write $\overline V(m,T)=V(\sigma_*(m,T);m,T)$. The equilibrium field satisfies the gap equation
\begin{equation}
 \Omega_\sigma=\frac{\sigma}{G}
 -4d\ST\frac{g(Q^2)M(Q^2)}{Q^2+M^2(Q^2)}=0.
 \label{eq:gap}
\end{equation}
The function $K$ is independent of temperature. We choose $K=0$ when quoting the subtracted condensate below. For pressure differences, one may equivalently subtract the zero-temperature equilibrium potential at the same parameters. This normalization leaves thermal quantities unchanged. Its mass dependence changes absolute source observables, so their background convention must be specified.

We take thermal derivatives at fixed spatial volume and fixed $m,G,\Lambda$, and kernel parameters, and mass-source derivatives at fixed $T,G,\Lambda$, and kernel. A vacuum refit changes these parameters and therefore defines a different operation. At $\sigma=0$, Eq.~\eqref{eq:V} vanishes, so Eq.~\eqref{eq:Bm} recovers the prescribed massive free gas.

\subsection{The thermal background and integrability}
Varying the same mass source in both flavors, we obtain the per-flavor condensate $C$ in the chosen subtraction convention and the entropy density $s$ from
\begin{equation}
 C=\frac{1}{N_f}\partial_m\Om,\qquad
 s=-\partial_T\Om.
\end{equation}
We decompose $s=\sint+\sfree$, with $\sint=-\partial_T\overline V$ and $\sfree=-\partial_T B_m$. Subscripts on the unminimized potential denote partial derivatives; all its other arguments are held fixed. At a scalar minimum with positive curvature, we use $\Omega_\sigma=0$ and differentiate the gap equation to obtain
\begin{align}
 \sigma_T&=-\frac{\Omega_{\sigma T}}{\Omega_{\sigma\sigma}},\\
 R\equiv\frac{C_V}{T}=\frac{\dd s}{\dd T}
 &=-\Omega_{TT}+\frac{\Omega_{\sigma T}^2}{\Omega_{\sigma\sigma}},
 \label{eq:schur}\\
 N_f\partial_TC&=-\partial_ms.
 \label{eq:maxwell}
\end{align}
Here $\sigma_T=\partial_T\sigma_*$ is the equilibrium field response. We first evaluate the partial derivatives of $\Omega$ and then insert the equilibrium solution. Positive scalar curvature makes the relaxation correction in Eq.~\eqref{eq:schur} nonnegative. We denote the fixed-field contribution by $\cfix=-\Omega_{TT}$.

Equation~\eqref{eq:gap} does not determine the thermal or source response: adding any smooth function of $T$ and $m$ leaves its $\sigma$ derivative unchanged. If all observables are derived consistently, Maxwell relations still hold. They therefore test a common generating potential without fixing its background.

For example, we can replace $B_m$ by $B_0=-7\pi^2dT^4/180$ without changing Eq.~\eqref{eq:gap}, since both backgrounds are independent of $\sigma$. The leading Stefan--Boltzmann term is also unchanged, because
\begin{equation}
 B_m=B_0+\frac{dm^2T^2}{12}
 +O\!\left(m^4\log\frac{T}{m}\right)
\end{equation}
at small $m/T$. For $m,T>0$, the source derivative nevertheless gives different condensates:
\begin{equation}
 C^{(B_m)}-C^{(B_0)}
 =4N_c\ip\frac{m}{E_m}n_F(E_m)>0,
 \label{eq:backgroundC}
\end{equation}
where $n_F(E)=(e^{E/T}+1)^{-1}$. At $m=0$, the background curvature difference is $N_f^{-1}\partial_m^2(B_m-B_0)=N_cT^2/6$. On a smooth equilibrium branch where the mass susceptibilities exist, this is also their difference. Thus agreement in the gap equation and leading high-temperature pressure is insufficient to determine the source response.

The role of the free term is especially clear if we combine $C_{\mathrm{int}}=N_f^{-1}\partial_m\overline V$ with the entropy derived from $\overline V+B_m$. For $m,T>0$, this inconsistent combination gives
\begin{align}
 N_f\partial_TC_{\mathrm{int}}+\partial_ms
 &=-\partial_T\partial_m B_m\\
 &=-\frac{4dm}{T^2}\ip n_F(E_m)[1-n_F(E_m)]<0.
\end{align}
The residual arises because one derivative omits the restored free term. We use this example to explain the requirement of a common potential; it does not describe an omission attributed to the nonlocal formulations cited above. All observables in the following sections are derived consistently from Eqs.~\eqref{eq:V} and \eqref{eq:Bm}. Their thermal signs remain a separate question.

\subsection{The statistical constraint on heat capacity}
\label{sec:canonical}
To interpret the heat capacity in Eq.~\eqref{eq:schur}, we compare it with the canonical variance relation. Consider a temperature-independent self-adjoint Hamiltonian $H$ on a Hilbert space with a positive-definite inner product, and assume that $Z=\operatorname{Tr}e^{-\beta H}$ and all required derivatives exist. At fixed spatial volume $\mathcal V$, with $\beta=1/T$ and $k_B=1$, we have
\begin{equation}
 C_V=\frac{1}{\mathcal V}\frac{\partial\langle H\rangle}{\partial T}
 =\frac{\langle H^2\rangle-\langle H\rangle^2}{\mathcal V T^2}\geq0.
 \label{eq:canonical-variance}
\end{equation}
At fixed chemical potential zero, the grand-canonical weight also reduces to $e^{-\beta H}$, so the same identity applies. A temperature-independent vacuum subtraction has no effect on this response.

A negative total $C_V$ therefore means that the computed potential cannot represent an exact equilibrium ensemble under these assumptions. Within that ensemble, it cannot be interpreted as an admissible equilibrium phase with negative heat capacity. Minimization in $\sigma$ tests the retained field direction, and $\Omega_{\sigma\sigma}>0$ alone does not establish the required temperature concavity of $\Om$. For a phenomenological action nonlocal in Euclidean time, convergent Euclidean integrals do not establish a Hamiltonian representation satisfying these assumptions. Moreover, the mean-field truncation need not preserve the exact statistical constraint. A failed sign test therefore rules out the specified approximation as an equilibrium EOS, without identifying whether the underlying problem comes from the spectrum or omitted dynamics. A negative fixed-field contribution by itself is less restrictive: the nonnegative relaxation term in Eq.~\eqref{eq:schur} may still give $C_V>0$.

\section{A controlled large-source entropy criterion}\label{sec:heavy}

\subsection{Assumptions and the scalar minimum}
We now study how the entropy behaves when the source becomes large at fixed couplings. We take $G,d,\Lambda>0$ and work on a closed, finite temperature interval $I$ whose lower endpoint is strictly above zero. We assume that $g$ is real, independent of $m$ and $\sigma$, and satisfies $0\leq g\leq1$, with a strictly positive combined Matsubara sum and momentum integral of $g$ at every $T\in I$. We also require $g$ to be continuously differentiable in $T$ and impose the convergence condition
\begin{equation}
 \sum_n\ip\sup_{T\in I}
 \left(1+\frac{Q^2}{\Lambda^2}\right)^2
 \left(g+T|g_T|\right)<\infty.
 \tag{H}\label{eq:H}
\end{equation}
Here $g_T$ is the temperature derivative at fixed spatial momentum and Matsubara index, including the dependence through $\omega_n$. The notation $\sup_{T\in I}$ means the largest value over the temperature interval. The assumptions and bound (H) must also hold on a slightly wider closed interval extending beyond both endpoints of $I$, with its lower endpoint still positive. This requirement allows us to differentiate both the kernel moments and the thermal remainder without assuming factorization of spatial and temporal dependence.

Under these assumptions, we find a unique global minimum over the real homogeneous scalar field for every $m>0$, with
\begin{equation}
 0<\sigma_*\leq\frac{4dGA}{m},\qquad
 V_{\sigma\sigma}(\sigma_*)>0,\qquad A=\ST g.
 \label{eq:sigmabound}
\end{equation}
Appendix~\ref{app:remainder} proves this statement over the retained real scalar field. Pion and vector fluctuations require separate stability tests, as do spatially modulated condensates, which have been studied in nonlocal models~\cite{Carlomagno2015}.

\subsection{The large-source expansion and entropy sign}
To express the expansion in terms of the kernel, we define two additional moments and a coefficient:
\begin{align}
 A_1&=\ST(Q^2g),\qquad H_g=\ST g^2,\\
 \mathcal B&=16d^2GAA_1+32d^3G^2A^2H_g.
 \label{eq:coeffB}
\end{align}
Under these assumptions, the large-$m$ expansion holds throughout $I$ with a common bound on each remainder:
\begin{align}
 \sigma_*={}&\frac{4dGA}{m}
 -\frac{4dG(A_1+4dGAH_g)}{m^3}+O_I(m^{-5}),
 \label{eq:sigmaexp}\\
 \overline V={}&-\frac{8d^2GA^2}{m^2}
 +\frac{\mathcal B}{m^4}+O_I(m^{-6}),
 \label{eq:Vexp}\\
 s={}&\sfree+\frac{\mathcal E_2}{m^2}
 -\frac{\mathcal B'}{m^4}+O_I(m^{-6}),
 \label{eq:sexp}\\
 \mathcal E_2={}&16d^2GAA'.
\end{align}
A prime denotes a temperature derivative. Each $O_I$ remainder has a common bound over $I$; neither that bound nor the large-mass threshold is asserted to be uniform as $T\to0$. Appendix~\ref{app:remainder} controls the thermal derivative of the determinant remainder directly. The coefficient dimensions are $[\mathcal E_2]=E^5$ and $[\mathcal B]=E^8$.

To determine the total entropy sign, we compare these interaction terms with the positive massive free entropy, which obeys
\begin{equation}
 \sfree\leq\frac{2dm^2T}{\pi^2}
 \left[4K_2(m/T)+\frac{m}{T}K_1(m/T)\right].
 \label{eq:freebound}
\end{equation}
Here $K_\nu$ is the modified Bessel function of the second kind of order $\nu$. We obtain this inequality from $\log(1+e^{-E_m/T})\leq e^{-E_m/T}$ and $n_F(E_m)\leq e^{-E_m/T}$. The modified-Bessel asymptotics~\cite{DLMFbessel} give a common upper bound that decreases exponentially with $m$ and applies to every temperature in $I$. It is therefore dominated by the algebraic interaction term whenever a displayed coefficient is nonzero.

Consequently, $A'(T_0)<0$ gives negative total entropy for all sufficiently large $m$ at fixed $T_0>0$, whereas $A'(T_0)>0$ gives positive entropy for all sufficiently large $m$. If $A'(T_0)=0$, a nonzero coefficient $-\mathcal B'(T_0)$ fixes the sign at order $m^{-4}$. The displayed terms leave the sign undetermined when both coefficients vanish. If $A'$ keeps a strict sign throughout the closed interval $I$, one sufficiently large mass threshold gives the same entropy sign at every temperature in that interval. The argument does not determine a rigorous numerical value for this threshold, and the error bounds need not remain valid as $T\to0$.

We can also use this expansion to determine what remains at the endpoint of a mass integration. Integrating Eq.~\eqref{eq:maxwell} between finite endpoints gives the exact relation
\begin{equation}
 s(T,m_0)=N_f\int_{m_0}^{m_{\max}}\partial_TC(T,m)\,\dd m
 +s(T,m_{\max}).
 \label{eq:reconstruction}
\end{equation}
This relation is the condensate reconstruction of Ref.~\cite{Isserstedt2021} for a model without an active gluonic contribution. With the prescribed massive-free completion, $s(T,\infty)=0$, although the approach may be from below with an algebraic tail. A calculation stopped at a finite mass therefore needs the corresponding tail correction. With the alternative $B_0$ completion, the mass-independent free entropy remains at the endpoint. The condensate integral alone cannot determine such a source-independent background.

\subsection{Positive Gaussian mixtures}
We first apply the criterion to a normalized radial mixture of Gaussians,
\begin{equation}
 g(Q^2)=\int_{(0,\infty)}e^{-uQ^2/\Lambda^2}\,\dd\rho(u),\quad
 \rho\geq0,\quad \int\dd\rho=1.
 \label{eq:mixture}
\end{equation}
Here $u>0$ is a dimensionless Gaussian width parameter and $\rho$ is a normalized nonnegative mixing measure. A sufficient condition for (H) is $\int(1+u^{-4})\,\dd\rho(u)<\infty$: the inverse fourth moment controls the largest momentum weight, while the nonzero Matsubara gap controls large $u$ on a compact interval at positive temperature. For each Gaussian component, the moment $A_u=\ST e^{-uQ^2/\Lambda^2}$ follows from Poisson summation:
\begin{equation}
 A_u(T)=\frac{\Lambda^4}{16\pi^2u^2}
 \theta_4\!\left(0,e^{-\Lambda^2/(4uT^2)}\right).
\end{equation}
For $0<z<1$, we use the Jacobi product~\cite{DLMFtheta}
\begin{equation}
 \theta_4(0,z)=\prod_{k=1}^\infty(1-z^{2k})(1-z^{2k-1})^2
\end{equation}
whose logarithmic derivative with respect to $z$ is strictly negative. Because $z$ increases with $T$, we have $A_u'<0$. The convergence condition permits taking the temperature derivative inside the mixing integral in Eq.~\eqref{eq:mixture}, giving $A'=\int A_u'\,\dd\rho<0$. Hence every kernel in this admissible class develops negative total entropy at sufficiently large source mass at each fixed positive temperature. This conclusion concerns source deformations of the prescribed mean-field functional at fixed $G$, $\Lambda$, and kernel; it does not assume that a light-quark calibration remains phenomenologically valid in the heavy-quark regime.

This result supplies a sufficient criterion and does not classify all kernels with negative tails. The positive mixture measure does not imply reflection positivity or a positive spectral representation of the dressed propagator. An instantaneous kernel independent of $\omega_n$ gives a divergent moment $A$, so it lies outside the assumptions. Equation~\eqref{eq:sexp} cannot be used to infer that limit by direct substitution.

\section{A covariant kernel with a sign-changing moment derivative}\label{sec:covariant}
To examine the role of the kernel shape, we introduce the dimensionless deformation parameter $\alpha$ and consider
\begin{equation}
 g_\alpha(Q^2)=e^{-y-\alpha y^2},\qquad
 y=Q^2/\Lambda^2,\qquad\alpha\geq0.
 \label{eq:quartic}
\end{equation}
Because $y^2$ contains the mixed term $2p^2\omega_n^2/\Lambda^4$, Eq.~\eqref{eq:quartic} deforms the covariant vacuum kernel radially rather than as independent spatial and temporal factors. Finite temperature still defines a thermal rest frame. For $\alpha>0$, the relation $\partial_y^2\log g_\alpha=-2\alpha<0$ rules out a positive exponential-mixture representation, since the logarithm of such a mixture has a nonnegative second derivative. We must nevertheless calculate the moment derivative to determine whether $A'$ becomes positive.

The sign is not fixed by covariance or radial monotonicity: Appendix~\ref{app:counterexample} gives a smooth positive radial counterexample with $A'>0$, together with the radial identities used in the calculation. For the quartic family, Eq.~\eqref{eq:threshold} supplies a sufficient high-temperature condition for $A'<0$.

\begin{figure*}[t]
\includegraphics[width=\textwidth]{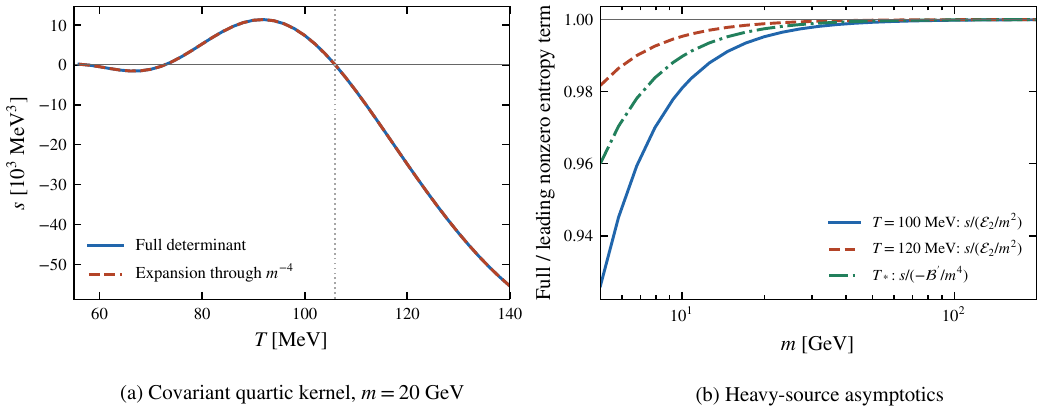}
\caption{Heavy-source entropy for the covariant $\alpha=1$ kernel, with the bare parameters of Table~\ref{tab:heavy} held fixed. (a) Full equilibrium entropy at $m=20\GeV$ and its expansion through $m^{-4}$; the dotted line marks the selected zero $T_*$ of $A'$. (b) Full entropy divided by its first nonzero asymptotic term at $T=100\MeV$, $120\MeV$, and $T_*$. At $T_*$ this term is $-\mathcal B'/m^4$. The massive free entropy is included and is exponentially small on these scales. Curves and points are numerical, not interval enclosures or physical pion refits.}
\label{fig:heavy}
\end{figure*}

At $\alpha=1$, $\Lambda=752.2\MeV$, and $G\Lambda^2=20.65$, a selected simple zero of $A'(T)$ occurs near $T_*=105.953\MeV$. At fixed $m=20\GeV$, the full entropy is positive at $100\MeV$ and negative at $120\MeV$; the free contribution is exponentially negligible. Figure~\ref{fig:heavy} compares the full result with the expansion. At the displayed temperatures away from $T_*$, its first nonzero interaction term is $\mathcal E_2/m^2$; at $T_*$ it is $-\mathcal B'/m^4$.

The entropy-zero temperature $T_0(m)$ consequently drifts as
\begin{align}
 T_0(m)&=T_*+\frac{\beta_{\rm drift}}{m^2}+O(m^{-4}),\\
 \beta_{\rm drift}&=\frac{\mathcal B'(T_*)}{\mathcal E_2'(T_*)}
 \simeq-6.65\times10^6\,\mathrm{MeV}^3.
 \label{eq:v2zerodrift}
\end{align}
The numerical checks in Appendix~\ref{app:counterexample} confirm this scaling, without supplying an interval proof or an exhaustive enumeration of zeros of $A'(T)$. These fixed-parameter calculations are separate from the light-source refits below.

\section{Vacuum matching and thermal-sign boundaries}\label{sec:fit}
\subsection{Light-source matching}
We now return to the light-source regime and refit the bare parameters. We use the bilocal currents of scheme II in Ref.~\cite{GomezDumm2006}, whose pion vertex is $g(q^2)$ at the average relative momentum. The pion pole, axial-current normalization, subtracted condensate, and pole-free continuation are specified in Appendix~\ref{app:enclosures}. This choice matters: the mass function alone does not determine the pion sector.

For the nominal fits, we use the pion mass $m_\pi$, pion decay constant $f_\pi$, and per-flavor subtracted vacuum condensate $C$ as exact computational inputs:
\begin{equation}
 m_\pi=139\MeV,\quad f_\pi=92.4\MeV,\quad -C=c^3.
 \label{eq:fitinputs}
\end{equation}
Here $c=(-C)^{1/3}$ is the positive mass scale used to specify the condensate input. At each $(\alpha,c)$, we refit $m,G,\Lambda$ and determine the vacuum field $\sigma_0\equiv\sigma_*(m,0)$ simultaneously. The condensate specified in Eq.~\eqref{eq:fitinputs} follows the subtraction in Eq.~\eqref{eq:vacC}, rather than the unsubtracted divergent expression. We treat these matching inputs as exact and assign no experimental input-error distribution.

\begin{figure*}[t]
\includegraphics[width=\textwidth]{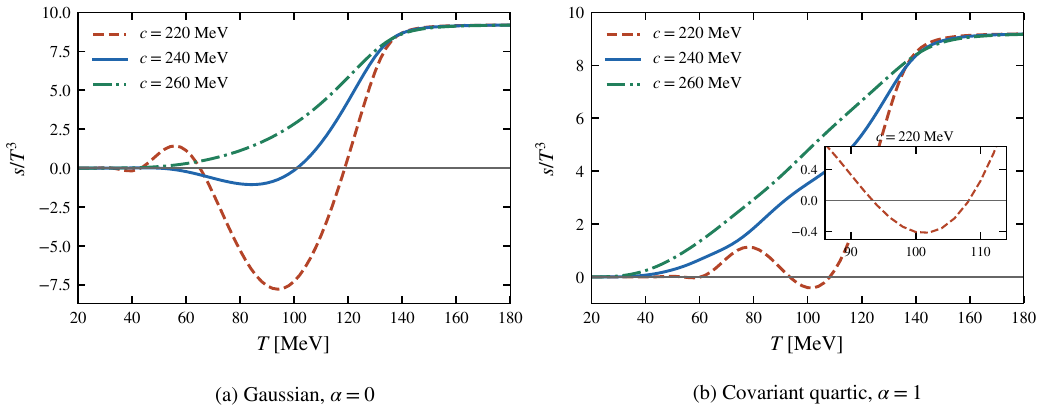}
\caption{Equilibrium entropy after refitting each kernel to $m_\pi=139\MeV$, $f_\pi=92.4\MeV$, and the per-flavor subtracted condensate $-C=c^3$. The Gaussian ($\alpha=0$, left) and quartic-exponent ($\alpha=1$, right) kernels use $c=220,240,260\MeV$. The inset resolves negative entropy for the quartic $c=220\MeV$ fit. The panel scales differ. All curves include the massive free contribution and connect numerical samples; they do not prove positivity between samples.}
\label{fig:fits}
\end{figure*}

\begin{table}[t]
\begin{ruledtabular}
\begin{tabular}{crrrr}
$\alpha$&$m$&$\Lambda$&$G\Lambda^2$&$\sigma_0$\\
\hline
0&5.7517&752.21&20.650&424.17\\
1&5.7609&1134.0&27.215&324.94
\end{tabular}
\end{ruledtabular}
\caption{Separate vacuum fits at $c=240\MeV$. The shared pion and condensate inputs are Eq.~\eqref{eq:fitinputs}. Masses and scales are in MeV. Rounded numerical fit values are shown; these rows are distinct from the interval-enclosed $c=225\MeV$ case below.}
\label{tab:fits}
\end{table}

\subsection{Kernel sensitivity at fixed vacuum observables}
Each kernel has its own refitted bare parameters (Table~\ref{tab:fits}). Figure~\ref{fig:fits} displays the entropy; the heat capacity is evaluated separately from Eq.~\eqref{eq:schur}. On the $T=20$--$300\MeV$ grid with $2\MeV$ spacing, both quantities become negative for the Gaussian $c=240\MeV$ and quartic $c=220\MeV$ fits, whereas neither has negative sampled values for the quartic $c=240\MeV$ or Gaussian $c=260\MeV$ fit. These comparisons establish sensitivity to both kernel and matching input; continuous positivity requires the bounds in Sec.~\ref{sec:certificate}.

\subsection{Entropy and heat-capacity folds}
We next follow the refitted $\alpha=1$ family and find that entropy zeros and zeros of its temperature slope disappear at different values of $c$. We call the merger of two temperatures at which the same thermal observable vanishes a fold. The entropy fold satisfies $s=s_T=0$, while the heat-capacity fold satisfies $R=\dd R/\dd T=0$, with $R=C_V/T$. Temperature derivatives include the response of the equilibrium thermal field and hold the fitted parameters fixed. Differentiation with respect to $c$ instead follows the vacuum-refitted parameter family.

\begin{figure*}[t]
\includegraphics[width=\textwidth]{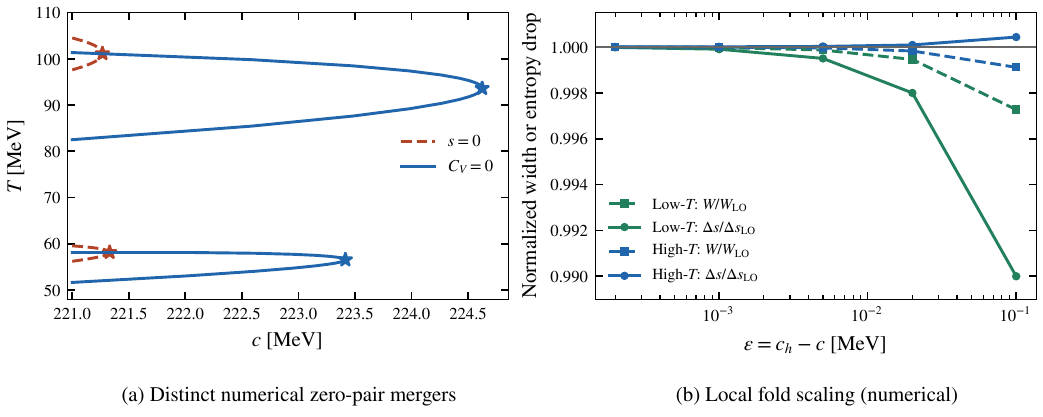}
\caption{Numerical continuation of four tracked zero-pair mergers of the refitted quartic-kernel family. (a) Entropy zeros (dashed) and heat-capacity zeros (solid); stars mark the numerical merger candidates. (b) Width $W=T_+-T_-$ and entropy difference $\Delta s$ of the two negative-heat-capacity intervals, normalized to $W_{\mathrm{LO}}=2\sqrt{2a_h/b_h}\sqrt\epsilon$ and $\Delta s_{\mathrm{LO}}=(4a_h/3)\sqrt{2a_h/b_h}\,\epsilon^{3/2}$, with $\epsilon=c_h-c$. The scaling tests Eq.~\eqref{eq:foldscaling}. Labels identify tracked components, without an exhaustive root enumeration. The plotted curves remain numerical; the local existence and uniqueness certificates are given in Appendices~\ref{app:stationary}--\ref{app:uniquefold}.}
\label{fig:folds}
\end{figure*}

\begin{table}[t]
\begin{ruledtabular}
\begin{tabular}{lrr}
Diagnostic&$c$ (MeV)&$T$ (MeV)\\
\hline
High-temperature entropy&221.267&101.125\\
Low-temperature entropy&221.331&58.138\\
Low-temperature heat capacity&223.415&56.536\\
High-temperature heat capacity&224.625&93.595
\end{tabular}
\end{ruledtabular}
\caption{Numerically continued thermal folds of the $\alpha=1$ family. These rounded coordinates remain numerical estimates; Appendices~\ref{app:stationary}--\ref{app:uniquefold} supply separate local certificates for all four folds. ``Low'' and ``high'' identify tracked components, not an exhaustive ordering of all roots.}
\label{tab:folds}
\end{table}

As Table~\ref{tab:folds} shows, negative entropy disappears before negative heat capacity along the tracked family. At both tracked heat-capacity mergers, the entropy is already positive, with approximate values $1.1459\times10^4$ and $9.0906\times10^5\,\mathrm{MeV}^3$, and the scalar curvature remains positive. The mergers therefore describe changes in thermal signs. They provide no evidence for a chiral phase transition or an exchange of scalar minima.

We check the numerical continuation through the local scaling near each heat-capacity fold. At $(c_h,T_h)$, define $a_h=\partial_cR>0$, $b_h=\partial_T^2R>0$, and $\epsilon=c_h-c$. The ordinary fold expansion then gives
\begin{align}
 R&=-a_h\epsilon+\tfrac12b_h(T-T_h)^2+\cdots,\\
 T_\pm&=T_h\pm\sqrt{2a_h/b_h}\,\sqrt\epsilon+O(\epsilon),\\
 \Delta s&\equiv s(T_-)-s(T_+)
 =-\int_{T_-}^{T_+}R\,\dd T\\
 &=\frac43a_h\sqrt{\frac{2a_h}{b_h}}\,\epsilon^{3/2}
 +o(\epsilon^{3/2}).
 \label{eq:foldscaling}
\end{align}
These are the exponents of an ordinary nondegenerate fold. Figure~\ref{fig:folds} tests the local scaling; numerical coefficient checks and the continuous proofs are given in Appendices~\ref{app:stationary} and \ref{app:uniquefold}. The four folds are each unique and nondegenerate within their separately specified local rectangles. This is not a global enumeration of thermal zeros.

\subsection{What is established beyond the numerical curves}\label{sec:stationaryzero}
For each tracked boundary, let $Y=s$ or $Y=R$. Along the true vacuum-refitted branch, we bound $Y$ at the condensate and temperature boundaries and establish $Y_c>0$ and $Y_{TT}>0$. The minimum of $Y$ over the temperature interval then rises through zero exactly once. At the resulting joint root,
\begin{equation}
 Y=Y_T=0,\qquad
 \det\frac{\partial(Y,Y_T)}{\partial(c,T)}=Y_cY_{TT}>0.
 \label{eq:v2foldcriterion}
\end{equation}
Thus two simple thermal zeros enclose a negative interval below each local fold, merge at the fold, and disappear above it. The proof propagates the vacuum-fit uncertainty and the thermal field response; it does not infer strict signs from numerical convergence. Appendices~\ref{app:stationary}--\ref{app:uniquefold} give the exact domains, sign margins, and derivative bounds.

\section{Positive thermal responses and an independent EOS test}\label{sec:certificate}
\subsection{A constructive example with positivity at every positive temperature}
We select $\alpha=1$ and the exact nominal inputs $(m_\pi,f_\pi,c)=(139,92.4,225)\MeV$ as a constructive example, without claiming that this fit is phenomenologically preferred or that the quartic kernel is generally more reliable than the Gaussian kernel. Appendix~\ref{app:enclosures} encloses the true vacuum solution in the parameter box of Table~\ref{tab:box}. Every thermal bound below holds throughout that computational enclosure; it does not represent an experimental input-error band.

Combining four continuous-temperature estimates gives
\begin{equation}
 s(T)>0,\qquad C_V(T)>0,\qquad T>0.
 \label{eq:v2allpositive}
\end{equation}
For $0<T\leq40\MeV$, the positive contribution of the isolated real pole to the fixed-field entropy slope exceeds the absolute contributions from the 26 retained complex-conjugate pole pairs and all remaining poles. A uniform bound also gives $s(0^+)=0$, so $ds/dT>0$ implies positive entropy. The comparison is relative: its dimensional lower bound vanishes exponentially as $T\to0$. A Taylor enclosure bridges $40$--$80\MeV$, adjoining interval cells cover $80$--$300\MeV$, and analytic free-gas and interaction bounds cover every $T\geq300\MeV$. All four calculations use the same parameter box and equilibrium branch. Their bounds are recorded in Appendix~\ref{app:enclosures}, with the complete low-temperature argument in Appendix~\ref{app:phase27-low-temperature}.

\begin{figure*}[t]
\includegraphics[width=\textwidth]{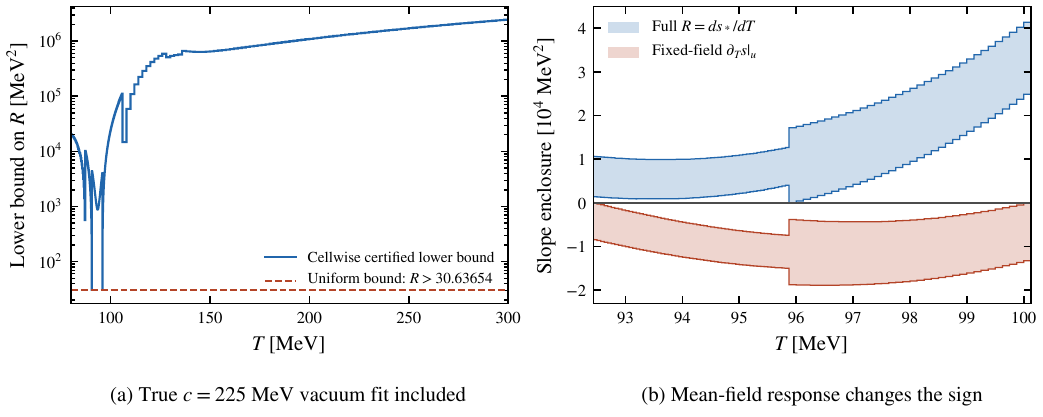}
\caption{Continuous-temperature enclosures for the $\alpha=1$, $c=225\MeV$ model, with the entire vacuum-fit parameter box propagated. Cell bounds establish positive $R=C_V/T$ on $80$--$300\MeV$. In the displayed subinterval the fixed-field entropy slope is strictly negative while the equilibrium slope remains strictly positive. Enclosure envelopes are conservative bounds, not physical uncertainty bands or estimates of the true minima.}
\label{fig:certificate}
\end{figure*}

Mean-field relaxation is essential to this result. Throughout $92.4375\leq T/\mathrm{MeV}\leq100.125$, the fixed-field slope is below $-22.68\,\mathrm{MeV}^2$, whereas the full $R=C_V/T$ exceeds $32.69\,\mathrm{MeV}^2$. The positive second term in Eq.~\eqref{eq:schur} therefore reverses the heat-capacity diagnosis (Fig.~\ref{fig:certificate}). The interval proofs include integration remainders and infinite tails and remain conditional on the reviewed integration implementation and interval engine, as detailed in Appendix~\ref{app:enclosures}.

\subsection{The EOS slope exceeds unity on a finite interval}\label{sec:eos-speed}
At zero chemical potential, the pressure $P=-\Om$ and energy density $\epsilon=-P+Ts$ give
\begin{equation}
 \frac{dP}{d\epsilon}=\frac{s}{C_V}=\frac{s}{TR}.
 \label{eq:v2eos}
\end{equation}
Positive $s$ and $C_V$ make $P$ and $\epsilon$ increasing functions of temperature, but do not constrain their slope ratio to be at most one. For the same nominal fit, direct bounds on the numerator and denominator establish
\begin{equation}
 \frac{dP}{d\epsilon}>1.047,
 \qquad 93.5\leq T/\mathrm{MeV}\leq93.5625.
 \label{eq:v2eosbound}
\end{equation}
A separate numerical evaluation at $93.5\MeV$ and the fit-box center gives approximately $1.93$. Appendix~\ref{app:eos-details} gives the uniform enclosure; the point estimate is not its lower bound, and the certified $0.0625\MeV$ interval is not the full width of the excess.

The nearby high-temperature heat-capacity fold explains how this enhancement arises. A uniform entropy bound on its certified rectangle establishes $s_h>0$ [Eq.~\eqref{eq:fold-entropy-positive}]. At fixed $T=T_h$, the ratio $s/(TR)$ therefore grows as $1/(c-c_h)$ as $c\to c_h^+$ along the refitted family. This reflects the vanishing heat capacity at an ordinary fold, without a new critical exponent or phase transition. The local expansion does not determine the enhancement at $c=225\MeV$; Eq.~\eqref{eq:v2eosbound} establishes that result independently.

To test the calibration sensitivity, we refit eight condensate inputs and numerically track $s/C_V$ over $80\leq T/\mathrm{MeV}\leq110$ (Table~\ref{tab:eos-input-scan}). The nominal fit has an above-unity interval of approximately $90.23$--$96.46\MeV$, of width $6.24\MeV$. The tracked peak decreases from $9.03$ at $c=224.7\MeV$ to $1.93$ at $225\MeV$ and $0.907$ at $225.5\MeV$; tracking this local peak to unity gives $c\simeq225.41\MeV$. These results show a calibration-sensitive enhancement near the heat-capacity fold. They are finite numerical scans, separate from the continuous certificate in Eq.~\eqref{eq:v2eosbound}.

The thermal momentum integrals already extend to infinity, unlike the restricted-integral examples of Refs.~\cite{Xue2022,Pasqualotto2023}. Full thermal integration and positive thermal responses therefore do not suffice for the usual EOS slope bound. For this temporally nonlocal action, the static derivative in Eq.~\eqref{eq:v2eos} does not determine a real-time front velocity or microscopic causality.

\subsection{Scope without thermal mesons}\label{sec:meson-scope}
Equation~\eqref{eq:v2allpositive} describes the quark mean field. Vacuum pion matching fixes a pole and an axial matrix element but adds no thermal pion population. Meson fluctuations beyond the scalar saddle generate these states~\cite{Hell2009}. Their formally subleading $1/N_c$ order~\cite{Radzhabov2011} does not prevent a lighter pion gas from dominating at low temperature.

At $T=93.5\MeV$, full Bose integrals for three ideal pions with fixed $m_\pi=139\MeV$ give $s_\pi/s\simeq0.82$ and $C_{V,\pi}/C_V\simeq5.6$. As a separate illustration, adding these ideal-gas values would give $(s+s_\pi)/(C_V+C_{V,\pi})\simeq0.53$, compared with the quark value $1.93$. The omitted contribution is therefore large enough to change the slope test in this temperature range. This sum includes neither scalar feedback nor vacuum refitting and is not a prediction of the meson-corrected model. Whether the folds and EOS excess persist requires a consistent fluctuation potential, its field and thermal derivatives, and matching at the same order. Appendix~\ref{app:meson-reference} gives the reference formulas and values. The mean-field excess is consequently a model diagnostic, not evidence for an anomalous sound speed in physical hadronic matter.

\section{Sensitivity to literature-motivated vacuum inputs}\label{sec:inputresponse}
\subsection{A deterministic domain and exact scale covariance}
We now vary the matching inputs to determine how strongly the exact-input result depends on the vacuum calibration. We choose the deterministic domain
\begin{align}
 m_\pi&\in[134.9763,139.57057]\MeV,\label{eq:pionmassdomain}\\
 f_\pi&\in[129/\sqrt2,131.4/\sqrt2]\MeV,\label{eq:piondecaydomain}\\
 c&\in[220,260]\MeV.
 \label{eq:inputdomain}
\end{align}
The pion-mass interval spans the neutral and charged PDG values, with errors rounded outward~\cite{PDG2025}; it is a declared interpolation between matching choices, not an error band for either species. The decay-constant range uses the PDG-preferred lattice value $130.2(1.2)\MeV$, including the charm-sea uncertainty, converted to our axial-current normalization~\cite{PDGDecay2025}. The condensate range covers representative inputs of the same separable-current construction~\cite{GomezDumm2006}; it is not a renormalized $\overline{\mathrm{MS}}$ error bar or evidence that all quartic-kernel fits are phenomenologically acceptable. We assign no likelihood, correlations, or joint confidence level to this deterministic domain; matching ambiguities in a statistical setting are discussed in Ref.~\cite{Pfaff2023}.

We refit every sampled input triple. Exact scale covariance allows us to vary the pion mass using the same dimensionless vacuum integrals. Defining $a=m_\pi/(139\MeV)$, we first fit the virtual inputs $(139\MeV,f_\pi/a,c/a)$ and then apply
\begin{equation}
 (m,\Lambda,\sigma,T)=a(m_v,\Lambda_v,\sigma_v,T_v),
 \qquad G=a^{-2}G_v.
 \label{eq:inputscaling}
\end{equation}
This transformation leaves the dimensionless coupling $\lambda=G\Lambda^2$ unchanged. Because the complete potential scales as $a^4$, the entropy $s$ and heat capacity $C_V$ scale as $a^3$, and the response $R=C_V/T$ scales as $a^2$. Under the same subtraction convention, these exact identities preserve both the vacuum and thermal gap equations.

\begin{figure*}[t]
\includegraphics[width=\textwidth]{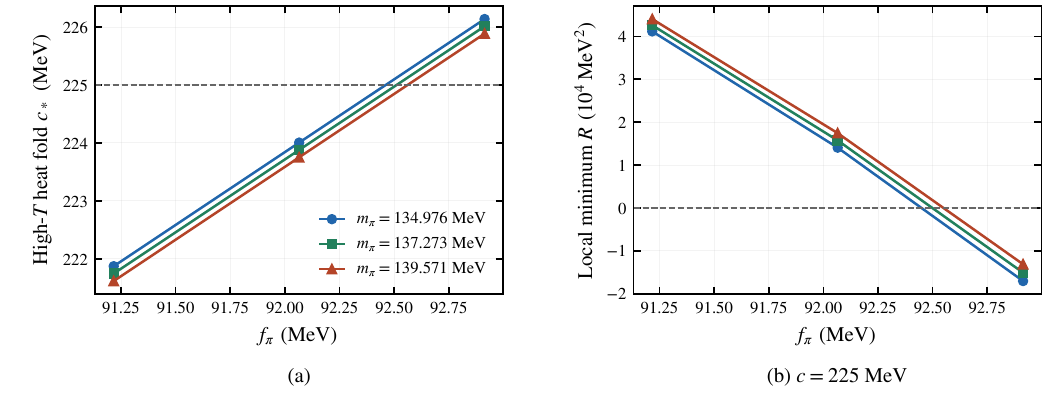}
\caption{Numerical dependence on refitted pion inputs: (a) the high-temperature heat-capacity fold and (b) its associated local minimum of $R$ at $c=225\MeV$. Markers denote the sampled pion-input pairs; connecting lines guide the eye and do not provide rigorous bounds between those pairs. Dashed lines mark $c_*=225\MeV$ and $R=0$.}
\label{fig:inputresponse}
\end{figure*}

\subsection{Refitted boundary shifts}
A $3\times3$ grid of pion-input endpoints and midpoints, with six condensate values per node, gives 54 refits and 36 tracked folds. Refined quadratures and independent refits check numerical convergence. Table~\ref{tab:inputfolds} reports sampled ranges, not bounds on the continuous domain.

\begin{table}[t]
\begin{ruledtabular}
\begin{tabular}{lcc}
Component&Sampled $c_*$ range&Sampled $T_*$ range\\
\hline
Low-$T$ $s$&218.364--222.824&57.377--58.493\\
High-$T$ $s$&218.300--222.764&99.820--101.706\\
Low-$T$ $C_V$&220.424--224.918&55.795--56.883\\
High-$T$ $C_V$&221.616--226.137&92.383--94.141
\end{tabular}
\end{ruledtabular}
\caption{Ranges across the nine numerical pion-input nodes. The lowest and highest sampled values are not rigorous bounds over the continuous domain in Eqs.~\eqref{eq:pionmassdomain}--\eqref{eq:inputdomain}. Both columns are in MeV.}
\label{tab:inputfolds}
\end{table}

The high-temperature heat-capacity fold moves across $c=225\MeV$ as the pion inputs vary (Fig.~\ref{fig:inputresponse}); three of the nine nodes have a negative local minimum of $R$ at that condensate input. Some continued entropy folds lie below $c=220\MeV$ and hence outside the declared domain. These finite-node observations assign neither signs to unsampled inputs nor probabilities to instability.

\subsection{A certified counterexample at fixed condensate input}
Using the exact scaling relation, we construct a fitted model with rigorously negative heat capacity inside the pion-input domain. Taking $c_-$ as the exact lower endpoint of the high-temperature heat-fold certificate, we choose
\begin{equation}
 a=225\MeV/c_-.
\end{equation}
We apply Eq.~\eqref{eq:inputscaling} to the strictly negative result in Table~\ref{tab:stationary}. The rescaled model satisfies the vacuum matching equations with exactly $c=225\MeV$; its inputs and the temperature at which heat capacity is negative are
\begin{align}
 m_\pi&\simeq139.23\MeV,\\
 f_\pi&\simeq92.554\MeV,\\
 T_w&\simeq93.751\MeV.
\end{align}
The exact scaled pion inputs both lie strictly inside the intervals in Eqs.~\eqref{eq:pionmassdomain} and \eqref{eq:piondecaydomain}. Rescaling the rigorous upper and lower bounds gives
\begin{align}
 -0.3539&<R(T_w)/\mathrm{MeV}^2<-0.2273,\\
 -33.18&<C_V(T_w)/\mathrm{MeV}^3<-21.31.
 \label{eq:inputcounterexample}
\end{align}
The proof uses the exact scaled inputs, not the displayed rounded decimals. At the same temperature, the nominal fit has $R>30.63\,\mathrm{MeV}^2$. Thus one rigorously negative example at exactly $c=225\MeV$ excludes uniform thermal positivity over the adopted pion-input domain. It does not classify the remaining inputs or attribute a sign reversal to pion experimental errors alone.

\section{Discussion and conclusions}\label{sec:conclusion}
We have studied thermodynamic consistency and stability in a covariant two-flavor nonlocal NJL model at zero chemical potential within the homogeneous quark mean-field approximation. At fixed couplings and kernel, a controlled large-source expansion gives a sufficient entropy-sign criterion in terms of Euclidean kernel moments. The leading interaction contribution is proportional to $AA'$, while the massive free entropy is exponentially suppressed. Every admissible positive Gaussian mixture therefore develops negative total entropy at sufficiently large source mass for each fixed $T>0$. This criterion requires no propagator-pole search and tests source deformations of the prescribed functional; it does not extrapolate a light-source vacuum fit to heavy-quark phenomenology.

In a separate light-source calculation, we refitted the model to pion observables and a subtracted condensate. The nominal fit with a quartic-exponent kernel has positive entropy and heat capacity for every $T>0$, including the response of the equilibrium mean field. Nevertheless, $dP/d\epsilon=s/C_V$ exceeds unity on a certified finite interval for the same fit. This constructive example shows that full thermal integration and positive thermal responses do not ensure the usual static EOS slope bound.

The fitted behavior depends on the matching inputs. Numerical refits show that the EOS slope enhancement decreases as the condensate input is increased over the sampled range. Locally near the heat-capacity fold, a vanishing positive heat capacity at finite positive entropy explains how large EOS slopes can arise. An exactly scaled, certified negative-heat-capacity counterexample further excludes uniform thermal positivity over the adopted matching-input domain. Thus the all-temperature positivity of the nominal fit cannot be extended to that entire domain.

These results distinguish common-potential consistency, stability in the retained scalar direction, thermal positivity, and the static EOS slope as separate diagnostics of the mean-field functional. They do not select a phenomenologically preferred kernel. Vacuum pion matching does not include thermal mesons, and the ideal-pion comparison shows that these omitted contributions can change the low-temperature slope test. Whether the mean-field excess survives consistent mesonic corrections remains unresolved. Moreover, the static derivative $dP/d\epsilon$ does not determine a real-time front velocity. Consistent mesonic fluctuations and real-time dynamics are needed to assess the corresponding hadronic thermodynamics and microscopic causality.

\begin{acknowledgments}
The work of X. W. was supported by the NSFC under Grants No. 12675170 and No. 12635010, and by the Anhui University of Science and Technology under Grant No. YJ20240001.
We thank ChatGPT (OpenAI) for assistance with code development and manuscript writing.
\end{acknowledgments}

\appendix
\section{Control of the thermal remainder}\label{app:remainder}
\subsection{Scalar minimum and large-source bound}
To prove Eq.~\eqref{eq:sigmabound}, we write $V_\sigma=\sigma/G-\mathcal R_\sigma$, where $\mathcal R_\sigma=4d\ST[gM/(Q^2+M^2)]$. For $\sigma>0$ and $g>0$, we obtain
\begin{equation}
 \partial_\sigma\log\frac{gM}{\sigma(Q^2+M^2)}
 =-\frac{m}{\sigma M}-\frac{2gM}{Q^2+M^2}<0.
\end{equation}
Thus $\mathcal R_\sigma/\sigma$ decreases strictly from infinity to zero, giving a unique positive gap root with positive curvature. Using $\mathcal R_\sigma\leq4dA/m$, we obtain the bound in Eq.~\eqref{eq:sigmabound}. For $a>0$, the inequalities $V(a)<V(-a)$ and $V(a)\geq a^2/(2G)-(4dA/m)a$ exclude a global minimum at negative $\sigma$ and show that $V$ grows without bound as $|\sigma|$ increases. The positive root is therefore the global minimum within the retained field space. Pion, vector, and spatially modulated directions lie outside this proof.

\subsection{Thermal remainder}
To control the large-source expansion, we set $\epsilon=m^{-2}$ and $v=m\sigma$. Equation~\eqref{eq:sigmabound} then bounds every equilibrium $v$ within the same closed, finite interval for all $T\in I$. We write the determinant logarithm as $L=\log(1+X)$, with
\begin{equation}
 a=2vg,\quad b=v^2g^2,\quad
 X=\frac{\epsilon a+\epsilon^2b}{1+\epsilon Q^2}.
\end{equation}
For sufficiently small $\epsilon$, we have $0\leq X\leq1/2$ uniformly. An expansion of the denominator in powers of $Q^2/m^2$ is not valid over the entire integration domain. To keep control of this region, we use the exact identities
\begin{align}
 X-\epsilon a-\epsilon^2(b-aQ^2)
 &=\epsilon^3\frac{aQ^4-bQ^2}{1+\epsilon Q^2},\\
 X-\epsilon a&=\epsilon^2\frac{b-aQ^2}{1+\epsilon Q^2},\\
 X^2-\epsilon^2a^2&=(X-\epsilon a)(X+\epsilon a).
\end{align}
The logarithmic remainder satisfies
\begin{align}
 \log(1+X)&=X-\tfrac12X^2+R_3(X),\\
 0\leq R_3&\leq X^3/3,\qquad \frac{\dd R_3}{\dd X}=\frac{X^2}{1+X}.
\end{align}
Consequently
\begin{equation}
 L=2\epsilon vg-\epsilon^2(2vQ^2g+v^2g^2)+r.
\end{equation}
Using $g^k\leq g$, $T|\partial_TQ^2|\leq2Q^2$, and $\epsilon Q^2/(1+\epsilon Q^2)\leq1$ in the exact remainder expressions gives
\begin{align}
 |r|+T|r_T|+\Lambda^2|r_v|
 \leq{}&C_I(\Lambda/m)^6
 \left(1+Q^2/\Lambda^2\right)^2\nonumber\\
 &\times(g+T|g_T|).
 \label{eq:remainderbound}
\end{align}
Here $C_I$ includes the fixed dimensional scales and the bounded range of $v$. Condition (H) makes the momentum integral and Matsubara sum of this bound finite and also controls the derivative of the outer Matsubara factor $T$.

In particular,
\begin{align}
 V(v/m;m,T)&=\epsilon f_0(v,T)+\epsilon^2f_1(v,T)+\mathcal R,\\
 f_0&=v^2/(2G)-4dAv,\\
 f_1&=4dA_1v+2dH_gv^2,
\end{align}
where $\mathcal R$, $\mathcal R_T$, and $\mathcal R_v$ are uniformly $O_I(m^{-6})$. Stationarity yields
\begin{align}
 v_*&=v_0+\epsilon v_1+O_I(\epsilon^2),\\
 v_0&=4dGA,\qquad v_1=-4dG(A_1+H_gv_0).
\end{align}
Because the leading term is stationary, the field displacement makes no linear contribution to the potential, giving Eq.~\eqref{eq:Vexp}. To obtain the entropy, we differentiate with respect to temperature at fixed field and then insert the stationary field, with the derivative error controlled by the remainder bound:
\begin{align}
 \sint={}&4d\epsilon A'v_*\nonumber\\
 &-\epsilon^2(4dA_1'v_*+2dH_g'v_*^2)+O_I(\epsilon^3).
\end{align}
The second coefficient is $4dA'v_1-4dA_1'v_0-2dH_g'v_0^2=-\mathcal B'$, which proves Eq.~\eqref{eq:sexp}. Our assumptions control one thermal derivative. Asymptotic results involving further derivatives, including the uniform implicit-root expansion in Eq.~\eqref{eq:zerodrift}, require integrable bounds on those additional derivatives throughout the temperature interval. The exponential family satisfies this stronger condition; we do not assume it for every kernel satisfying only (H).

\section{Kernel-sign constructions and numerical checks}\label{app:counterexample}

\subsection{Radial identities and the quartic sign bound}
For a sufficiently smooth and decaying radial function $f(Q^2)$ whose temperature dependence enters only through the Matsubara frequencies, integration by parts yields
\begin{equation}
 \partial_T\ST f
 =\frac{1}{\pi^2}\sum_{n\geq0}\int_0^\infty
 (p^2-\omega_n^2)f\,\dd p.
 \label{eq:radial}
\end{equation}
The radial measure in this expression contains no additional $p^2$. Applying it to the logarithm in Eq.~\eqref{eq:V}, we obtain a second way to calculate the interaction entropy:
\begin{equation}
 \sint=\frac{2d}{\pi^2}\sum_{n\geq0}\int_0^\infty
 (p^2-\omega_n^2)\log\frac{Q^2+M^2}{Q^2+m^2}\,\dd p.
 \label{eq:radials}
\end{equation}
We differentiate at fixed field and then set it to $\sigma_*$. When the upper limit is a finite $p_{\max}$, the surface term $\pi^{-2}\sum_{n\geq0}\omega_n^2p_{\max} f(p_{\max}^2+\omega_n^2)$ must be included in Eq.~\eqref{eq:radial}. We retain this term in the numerical cross-check.

Using $x=\omega_n^2/\Lambda^2$, we can also differentiate the moment directly:
\begin{equation}
 A'=\frac{1}{\pi^2}\sum_{n\geq0}\int_0^\infty
 p^2g_\alpha[1-2x(1+2\alpha y)]\,\dd p.
\end{equation}
Since $y\geq x\geq\pi^2T^2/\Lambda^2$, this expression gives the sufficient condition
\begin{equation}
 T\geq\frac{\Lambda}{\pi\sqrt{1+\sqrt{1+4\alpha}}}
 \quad\Longrightarrow\quad A'<0.
 \label{eq:threshold}
\end{equation}
The condition bounds the region of negative derivative without locating a zero of $A'(T)$. The next subsection constructs an analytic smooth radial example with $A'>0$. Thus covariance, nonnegativity, and monotonicity of $g$ do not by themselves determine the sign.

\subsection{A smooth positive-sign construction}

For an explicit sign calculation, we first consider the radial cutoff with momentum scale $Q_c>0$, $g_{Q_c}(Q^2)=\mathbf1_{Q^2<Q_c^2}$, which equals 1 for \mbox{$Q^2<Q_c^2$} and 0 otherwise. In $Q_c/(3\pi)<T<Q_c/\pi$, only the lowest Matsubara pair contributes, giving
\begin{align}
 A_{Q_c}&=\frac{T}{3\pi^2}(Q_c^2-\pi^2T^2)^{3/2},\\
 A_{Q_c}'&=\frac{\sqrt{Q_c^2-\pi^2T^2}}{3\pi^2}(Q_c^2-4\pi^2T^2).
\end{align}
Thus $A_{Q_c}'>0$ on $Q_c/(3\pi)<T<Q_c/(2\pi)$. To obtain a kernel to which the differentiable theorem applies, we choose a closed temperature interval strictly inside this positive range and approximate the ball by bounded smooth monotone radial cutoffs. The interval stays a positive distance from temperatures where Matsubara frequencies cross the ball boundary, so dominated convergence in Eq.~\eqref{eq:radial} gives convergence of $A'$ there. We then mix in a Gaussian kernel with arbitrarily small positive weight, keeping the total weight equal to one. The resulting kernel is strictly positive everywhere, retains its normalization and monotonicity, satisfies (H), and still has $A'>0$. Radial covariance and positivity therefore do not imply a universally negative sign.

\subsection{Large-source numerical checks}
For the fixed bare parameters in Table~\ref{tab:heavy}, the selected numerical zero of $A'(T)$ is $T_*\simeq105.953\MeV$; the sufficient negative-sign threshold in Eq.~\eqref{eq:threshold} is $133.10\MeV$. The table and Fig.~\ref{fig:heavy} test the large-source expansion, separately from the light-source vacuum refits.

\begin{table}[!htbp]
\begin{ruledtabular}
\begin{tabular}{crr}
$T$ (MeV)&$\sint$&$\delta_{\mathrm{NLO}}$\\
\hline
100&7343.5&$1.07\times10^{-5}$\\
$T_*$&$-24.717$&$2.57\times10^{-3}$\\
120&$-24873$&$1.93\times10^{-6}$
\end{tabular}
\end{ruledtabular}
\caption{Heavy-source illustration at $m=20\GeV$, $\alpha=1$, $\Lambda=752.2\MeV$, and $G\Lambda^2=20.65$. Entropies are in $\mathrm{MeV}^3$. NLO includes both algebraic terms in Eq.~\eqref{eq:sexp}; the relative difference $\delta_{\mathrm{NLO}}=|s_{\mathrm{int}}-s_{\mathrm{int}}^{\mathrm{NLO}}|/|s_{\mathrm{int}}|$ is evaluated before rounding. The free-entropy upper bound is below $10^{-60}\,\mathrm{MeV}^3$ at all three points. The signs are converged numerical results.}
\label{tab:heavy}
\end{table}

The finite $m^{-4}$ contribution controls the entropy at $T_*$.

For this exponential kernel, bounds on the additional derivatives allow us to follow the entropy zero as the mass changes. Expanding the equation for this zero gives
\begin{align}
 T_0(m)&=T_*+\frac{\beta_{\rm drift}}{m^2}+O(m^{-4}),
 \label{eq:zerodrift}\\
 \beta_{\rm drift}&=\frac{\mathcal B'(T_*)}{\mathcal E_2'(T_*)}
 \simeq-6.6485\times10^6\,\mathrm{MeV}^3.
\end{align}
The numerical root shifts agree with Eq.~\eqref{eq:zerodrift} through the expected $m^{-4}$ residual. Independent high-precision quadratures check convergence; they are not interval certificates. Detailed precision comparisons are retained in the accompanying computational records.

\section{Vacuum matching, thermal bounds, and EOS diagnostics}\label{app:enclosures}
We use $\mu_m=m/\Lambda$, $u_0=\sigma_0/\Lambda$, and $\lambda=G\Lambda^2$ for the dimensionless vacuum parameters. The exact nominal matching inputs are held fixed; the intervals below enclose computational errors rather than experimental uncertainties.

\subsection{Pion observables and axial normalization}
We first specify the pion and axial vertices needed for the vacuum fit, because the mass function alone leaves this part of a nonlocal model undetermined. We adopt the bilocal-current construction of scheme II in Ref.~\cite{GomezDumm2006},
\begin{equation}
 j_a(x)=\int\dd^4z\,g(z)\,\bar\psi(x+z/2)\Gamma_a\psi(x-z/2),
\end{equation}
where $\psi$ is the two-flavor quark field, $\bm\tau$ are Pauli matrices in flavor space, and $\Gamma_a=(1,i\gamma_5\bm\tau)$ labels the scalar and three pseudoscalar channels. The coordinate-space function $g(z)$ is the Fourier transform of the momentum-space form factor $g(q^2)$. The interaction uses the scalar--pseudoscalar coupling $G$. In momentum space, this construction gives the pion vertex $g(q^2)$ at the average momentum. Using a product of separate leg form factors would define a different construction for the vacuum fit.

At zero temperature, we introduce $\iq=\int\dd^4q/(2\pi)^4$, $M=m+\sigma_0g$, and $D=q^2+M^2$. We determine the scalar field and the per-flavor subtracted condensate from
\begin{align}
 \frac{\sigma_0}{G}&=8N_c\iq\frac{gM}{D},\\
 C&=-4N_c\iq\left(\frac{M}{D}-\frac{m}{q^2+m^2}\right).
 \label{eq:vacC}
\end{align}

\begin{table}[!htbp]
\begin{ruledtabular}
\begin{tabular}{cl}
Parameter&Enclosure\\
\hline
$\mu_m$&$[0.00699129489800533,$\\
&$\phantom{[}0.00699129489802031]$\\
$\Lambda/\mathrm{MeV}$&$[989.29777783165,\ 989.29777783271]$\\
$\lambda$&$[31.167472074415,\ 31.167472074634]$
\end{tabular}
\end{ruledtabular}
\caption{Outward-rounded intervals bounding the $\alpha=1$, $c=225\MeV$ fitted parameters used in the thermal proof. The proofs use the original exact rational endpoints in the accompanying data; the wider displayed intervals are only for presentation. Each bracket bounds a parameter and is not a statistical error bar.}
\label{tab:box}
\end{table}
For Euclidean external pion momentum $P$, we use $q_\pm=q\pm P/2$, $M_\pm=M(q_\pm^2)$, $D_\pm=q_\pm^2+M_\pm^2$, and $\mathcal N=q_+\cdot q_-+M_+M_-$. The two loop functions entering the pion sector are
\begin{align}
 J(P^2)&=8N_c\iq g(q^2)\frac{\mathcal N}{D_+D_-},\\
 K(P^2)&=8N_c\iq g^2(q^2)\frac{\mathcal N}{D_+D_-}.
\end{align}
These loops determine the inverse pion propagator, its residue, and the decay constant through
\begin{align}
 G_\pi(P^2)&=G^{-1}-K(P^2),\quad G_\pi(-m_\pi^2)=0,\\
 Z_\pi^{-1}&=-K'(-m_\pi^2)>0,\\
 m_\pi^2f_\pi&=m\sqrt{Z_\pi}J(-m_\pi^2).
 \label{eq:pion}
\end{align}
Here $G_\pi$ is the inverse pion propagator, $Z_\pi$ is its propagator pole residue, and the prime denotes a $P^2$ derivative. We obtain the last relation from the gauged nonlocal axial current. The equivalent form follows by introducing $\mathcal F=mJ+\sigma_0K$, which gives $f_\pi=\sqrt{Z_\pi}[\mathcal F(-m_\pi^2)-\mathcal F(0)]/m_\pi^2$.

We check the implementation against the Ward identity $\sigma_0G_\pi(0)=mJ(0)$, the chiral Goldberger--Treiman relation $f_{\pi,\chi}\sqrt{Z_{\pi,\chi}}=\sigma_{0,\chi}$, and the leading Gell-Mann--Oakes--Renner relation $m_\pi^2f_{\pi,\chi}^2=-2mC_\chi+O(m^2)$. The subscript $\chi$ denotes the zero-temperature chiral limit $m\to0$ at fixed coupling and kernel; in particular, $\sigma_{0,\chi}$ is the vacuum field in that limit. Because these checks share the same loop ingredients, they test the structure of the calculation without providing an independent implementation of the full axial vertex. We also exclude poles in the strip required for continuation to the pion point, as described below.

\subsection{Vacuum matching equations and pole-free pion strip}
For $\alpha=1$, let $r=|q|/\Lambda$, $g=e^{-r^2-r^4}$, and $\widehat M=\mu_m+u_0g$. The dimensionless vacuum integrals
\begin{align}
 A_v&=\frac3{\pi^2}\int_0^\infty
 r^3\frac{g\widehat M}{r^2+\widehat M^2}\,\dd r,\\
 B_v&=\frac3{2\pi^2}\int_0^\infty
 r^3\frac{u_0g(r^2-\mu_m\widehat M)}
 {(r^2+\widehat M^2)(r^2+\mu_m^2)}\,\dd r
\end{align}
satisfy $\lambda=u_0/A_v$ and $-C=\Lambda^3B_v$. With $\widehat J=J/\Lambda^2$, $\widehat K=K/\Lambda^2$, and $\widehat K'$ differentiated with respect to $P^2/\Lambda^2$, the three residuals are
\begin{equation}
 \bm F_v=\begin{pmatrix}
 1-u_0\widehat K/A_v\\[2pt]
 \displaystyle\frac{\mu_m\Lambda^3\widehat J}
 {m_\pi^2f_\pi\sqrt{-\widehat K'}}-1\\[5pt]
 \Lambda^3B_v/c^3-1
 \end{pmatrix}.
 \label{eq:vacresidual}
\end{equation}
We evaluate the pion quantities at $P^2=-m_\pi^2$. The hatted loop functions also depend on the corresponding dimensionless external momentum.

The pion-loop continuation requires a strip of complex energy free of propagator poles for every integration momentum; checking sampled momenta does not suffice. For the family in Eq.~\eqref{eq:quartic}, we obtain a sufficient criterion by setting $a=m_\pi/(2\Lambda)$, $b=1-6\alpha a^2>0$, and $E=e^{a^2}$. Define
\begin{equation}
 R_{\max}=\begin{cases}
 1,&b\geq2\alpha,\\
 (2\alpha/b)e^{-1+b/(2\alpha)},&0<b<2\alpha.
 \end{cases}
\end{equation}
If
\begin{equation}
 2u_0ER_{\max}(\mu_m+u_0E)<1,\qquad \mu_m+u_0>a,
 \label{eq:strip}
\end{equation}
the denominator has no zeros in the required strip $|\operatorname{Im}q_4|/\Lambda\leq a$. We verify these inequalities for every combination of fitted parameters within their allowed intervals. To show why they exclude poles, write the dimensionless energy as $x+iv$ and set $t=\bm r^2+x^2$, $z=t-v^2+2ixv$, and $\xi=2xv$. The kernel has the form $g(z)=A_z e^{-i\xi c_z}$, with $c_z=1+2\alpha(t-v^2)>0$, $A_z\leq E e^{-\alpha t^2-bt}$, and $A_zc_z\leq ER_{\max}$. For $\xi\neq0$, its dimensionless denominator obeys
\begin{equation}
 \frac{\operatorname{Im}[z+(\mu_m+u_0g(z))^2]}{\xi}
 \geq1-2u_0ER_{\max}(\mu_m+u_0E)>0.
\end{equation}
At $v=0$ the denominator is positive and real. At $x=0$, any potentially negative $z$ lies in $[-a^2,0]$. Since $b>0$ implies $\alpha a^2<1/6$, $g(z)\geq1$ there and the denominator is at least $(\mu_m+u_0)^2-a^2>0$. This argument excludes poles for every spatial momentum within the pion strip. It does not extend the exclusion to the whole complex plane.

Interval automatic differentiation encloses the vacuum loops and their Jacobian. Finite radial and angular quadrature errors are bounded by Cauchy estimates, all infinite tails are retained, and $-\widehat K'>0$ is checked before taking the square root.

Let $x=(\mu_m,u_0,\Lambda)^\mathsf{T}$ be the vacuum-fit variables, $X$ their parameter box, $x_0$ its center, $J_v(X)$ an interval matrix bounding the residual Jacobian throughout $X$, and $C_0$ a rational approximate inverse of the Jacobian at $x_0$. In the following expression, $I$ is the identity matrix. The Krawczyk condition~\cite{Krawczyk1969,Moore2009}
\begin{align}
 \mathcal K(X)&=x_0-C_0\bm F_v(x_0)
 +(I-C_0J_v(X))(X-x_0),\\
 \mathcal K(X)&\subset\operatorname{int}X
\end{align}
together with the contraction bound establishes exactly one vacuum matching solution in $X$. Mean-value intersections refine this enclosure while retaining the root. The contraction factor is below $0.005523$. All thermal bounds use the complete exact parameter box underlying Table~\ref{tab:box}, including points away from the vacuum constraints; the matching inputs remain fixed. This propagates computational fitting uncertainty, not experimental errors.

\subsection{Equilibrium response and mean-field relaxation}

To calculate the equilibrium thermal response, we set $u=\sigma/\Lambda$, use $r=|\bm p|/\Lambda$ for the dimensionless spatial momentum, and define the dimensionless gap residual $F$ by
\begin{equation}
 F(u,T)=\frac{u}{\lambda}
 -\frac{24T}{\Lambda\pi^2}\sum_{n\geq0}\int_0^\infty
 r^2\frac{g\widehat M}{q+\widehat M^2}\,\dd r,
 \label{eq:thermalF}
\end{equation}
where $q=r^2+[(2n+1)\pi T/\Lambda]^2$ and $\widehat M=\mu_m+ug(q)$. Since $\partial_u\Omega=\Lambda^4F$, the response in Eq.~\eqref{eq:schur} takes the form
\begin{equation}
 R=\cfix+\Lambda^4\frac{F_T^2}{F_u}.
 \label{eq:dimensionlessR}
\end{equation}
Gap endpoint signs and a positive lower bound on $F_u$ enclose the equilibrium field. The continuous bounds on $80\leq T\leq300\MeV$ are

\begin{align}
 s&>6.491\times10^5\,\mathrm{MeV}^3,\\
 R=C_V/T&>30.63\,\mathrm{MeV}^2,\\
 C_V&>2.450\times10^3\,\mathrm{MeV}^3.
 \label{eq:positivebounds}
\end{align}
The bounds cover the entire parameter box and need not equal the physical minima. On

\begin{equation}
 92.4375\leq T\leq100.125\MeV,
\end{equation}
we obtain the simultaneous bounds
\begin{align}
 \cfix&<-22.68\,\mathrm{MeV}^2,\\
 R&>32.69\,\mathrm{MeV}^2.
 \label{eq:relaxationbounds}
\end{align}
These simultaneous bounds quantify the relaxation effect discussed in Sec.~\ref{sec:certificate}.

\subsection{Temperatures of 300 MeV and above}\label{sec:hightemperature}
The analytic estimates below give, throughout the parameter box and every $T\geq300\MeV$,

\begin{align}
 \frac{s}{T^3}&>9.207, &
 \frac{C_V}{T^3}&>26.99,
 \label{eq:highpositive}
\end{align}
uniformly over the full parameter box. The separate free and interaction bounds are derived in Appendix~\ref{app:high-temperature}, without differentiating an entropy inequality.

\subsection{Low temperatures and completion of the positivity proof}
\label{sec:low-positive}
Appendix~\ref{app:phase27-low-temperature} encloses the thermal field and the complete pole series on $0<T\leq40\MeV$. Write $R_{\rm fixed}=\cfix$ for the fixed-field entropy slope, and $R_{\rm other}$ for the sum of all pole contributions except that of the isolated real pole. With $L(T)>0$ a lower bound on the real-pole contribution and $\rho$ a uniform upper bound on the ratio of the absolute remaining contributions to $L(T)$, it establishes

\begin{align}
 |R_{\rm other}(T)|&\leq\rho L(T),\\
 \rho&<0.900,\\
 R(T)&\geq R_{\rm fixed}(T)\geq(1-\rho)L(T)>0.
 \label{eq:low-positive}
\end{align}
Here $L(T)$ vanishes exponentially as $T\to0$; no positive temperature-independent dimensional lower bound is implied. The same appendix proves $s(0^+)=0$, so $R>0$ implies $s>0$. Taylor enclosures bridge $40$--$80\MeV$ and give

\begin{align}
 s&>595.2\,\mathrm{MeV}^3,\\
 R&>33.35\,\mathrm{MeV}^2,\\
 C_V&>1334\,\mathrm{MeV}^3.
 \label{eq:bridge-positive}
\end{align}
The four bounds cover every $T>0$ on the same equilibrium branch and exact parameter box.

\subsection{Finite thermal integrals and tails}
For each temperature subinterval in $80$--$300\MeV$, we bound the finite radial integrals on $0\leq r\leq4$ by integrating a degree-seven Taylor expansion with interval coefficients and bounding the remainder with the eighth derivative. For the response, we retain six positive frequencies below $100\MeV$ and four otherwise, and bound all remaining frequencies analytically. Before evaluating Eq.~\eqref{eq:dimensionlessR}, we verify that the gap residual changes sign across the field interval and that the curvature is positive throughout it. Both the fixed-field slope and the gap response include the kernel derivatives.

After bounding the prefactors for each observable, we reduce the tail estimate to a bound on the sum and integral $\sum_{n\geq0}\int_0^\infty e^{-(q+q^2)}\,\dd r$. On a temperature cell $[T_a,T_b]$, let $\Lambda_{\max}$ be the upper endpoint of the fitted scale interval and set $\tau_{\min}=T_a/\Lambda_{\max}$, $v_n=[(2n+1)\pi\tau_{\min}]^2$, and $h_n=v_n+v_n^2$. If $P$ is the radial cutoff and $N$ the number of retained positive frequencies, the omitted part of this measure is bounded by $K_r+K_n$, where
\begin{align}
 K_r&=\frac{e^{-(P^2+P^4)}}{2P+4P^3}
 \sum_{n<N}e^{-h_n},\\
 K_n&=\frac{\sqrt\pi}{2}\frac{e^{-h_N}}
 {1-e^{-(h_{N+1}-h_N)}}.
\end{align}
These bounds use the full decay factor $e^{-(q+q^2)}$. Polynomial factors are absorbed into a bound proportional to the more slowly decaying kernel $e^{-(q+q^2)/2}$; its radial denominator is $P+2P^3$ and its frequency prefactor is $\sqrt{\pi/2}$. All amplitudes and dimensional prefactors use bounds valid throughout the fitted-parameter domain. The resulting bounds control the omitted gap, entropy, curvature, and response integrals. For the free thermal component, monotonicity and explicit integral remainders give an enclosure at the extreme masses and temperatures. This treats the free term and every infinite momentum and frequency tail explicitly; their bounds do not follow from finite samples.

\subsection{High-temperature upper bounds}\label{app:high-temperature}
For the same enclosed $c=225\MeV$ fit, we can bound the infinite Matsubara sum directly at every $T\geq T_0=300\MeV$. Write $t=T/\Lambda$, $v_n=[(2n+1)\pi t]^2$, and $h_n=v_n+v_n^2$. The quartic kernel satisfies
\begin{align}
 g(r^2+v_n)^k&\leq e^{-k(r^2+r^4)}e^{-kh_n},\\
 \sum_{n\geq0}e^{-kh_n}&\leq
 \overline S_k(t)=\frac{e^{-kh_0}}{1-e^{-k(h_1-h_0)}}.
 \label{eq:highsum}
\end{align}
The frequency increments increase with $n$, and $h_1-h_0=8v_0+80v_0^2$. Both factors in the upper bound decrease as $T$ increases. Using Gaussian radial integrals and the dimensionless denominator bound $q+\widehat M^2\geq\pi^2t^2$, we obtain
\begin{equation}
 a_k=\frac{6\lambda\overline S_k(t)}{\pi^{7/2}t k^{3/2}},
 \qquad u\leq\frac{\mu_m a_1}{1-a_2}\quad(a_2<1).
\end{equation}
For a model parameter, a subscript $+$ or $-$ denotes the upper or lower endpoint of its enclosing interval. Let $a_k^+=a_k(t_0,\lambda_+)$, with $t_0=T_0/\Lambda_+$, and define $U=\mu_{m,+}a_1^+/(1-a_2^+)$. At the endpoint, we find $a_2^+<0.12437$ and $0<u\leq U<0.015863$. The curvature also obeys $F_u\geq(1-a_2)/\lambda>0.028094$ on the entire real scalar axis throughout this temperature range.

Put $B=\mu_{m,+}+U$ and $\delta=U(2\mu_{m,+}+U)$. Integrating the entropy by parts over the complete radial domain, and bounding its fixed-field derivative separately, yields
\begin{align}
 \frac{|s_{\rm int}|}{T^3}
 &\leq \frac{12\Lambda_+^3}{\pi^2T_0^3}
       \delta\frac{\sqrt\pi}{2}\overline S_1(t_0),\\
 \frac{|R_{\rm int,fix}|}{T^2}
 &\leq\frac{12\Lambda_+^3}{\pi^2T_0^3}
       (4\delta+72UB)\sqrt{\frac\pi2}\overline S_{1/2}(t_0).
 \label{eq:highmajorants}
\end{align}
Here $R_{\rm int,fix}=-\partial_T^2V$ at fixed field, and $R_{\rm free}=\partial_T\sfree$. For example, the derivative integrand is bounded by
$[4\delta+4UBq+8UBq^2]g\leq(4\delta+72UB)\sqrt g$.
The second line therefore bounds the derivative integrand directly; differentiating the first bound would not give it. The free lower bounds
$s_{\rm free}/T^3\geq7\pi^2d/45-dm^2/(6T^2)$ and
$R_{\rm free}/T^2\geq7\pi^2d/15-dm^2/(3T^2)$ hold at all positive masses and temperatures. Combining them with the nonnegative response in Eq.~\eqref{eq:schur} proves Eq.~\eqref{eq:highpositive} for the complete parameter box.

The elementary bounds are also checked with rational series and integer square-root brackets.

\subsection{Finite-window construction and numerical controls}
The middle-temperature construction uses 318 adjoining cells on $80$--$300\MeV$; 89 cells supply the fixed-field comparison. Exact cell endpoints and quadrature settings are retained in the certificates.

For $40\leq T\leq80\MeV$, we use 80 adjoining temperature intervals and bound the equilibrium slope on each interval by expanding about its center,
\begin{equation}
 R(T)\in R(T_c)+R_T(T_c)[-h,h]
       +R_{TT}(J)[0,h^2/2],
 \label{eq:bridge-taylor}
\end{equation}
where $T_c$ is the cell center, $h$ its half-width, and $J=[T_c-h,T_c+h]$. We bound the center values and the second derivative throughout each temperature interval, including the full fitted-parameter and equilibrium-field intervals, interval quadrature remainders, and analytic radial and frequency tails. A directly computed positive entropy bound at $40\MeV$ supplies the starting value. An independent arithmetic calculation verifies that the cells leave no gaps and reproduces the bounds in Eq.~\eqref{eq:bridge-positive}.

High-precision potential derivatives, pole sums, EOS values, and refitted fold locations provide separate convergence checks. Their detailed tolerances and sampled values are retained in the computational records; none replaces a continuous interval bound.

\begin{table}[!htbp]
\begin{ruledtabular}
\begin{tabular}{rrrr}
$c$ (MeV)&Peak $s/C_V$&$T$ range (MeV)&Width (MeV)\\
\hline
224.7&9.03&89.08--97.29&8.21\\
225&1.93&90.23--96.46&6.24\\
225.25&1.21&91.54--95.41&3.87\\
225.5&0.907&---&---\\
226&0.628&---&---\\
228&0.340&---&---\\
230&0.268&---&---\\
240&0.204&---&---
\end{tabular}
\end{ruledtabular}
\caption{Numerically refitted EOS sensitivity on $80\leq T/\mathrm{MeV}\leq110$. The temperature range is the tracked interval with $s/C_V>1$; a dash means none was found in this window. Peaks, crossing temperatures, and widths are numerical estimates, not interval bounds. Widths use the unrounded crossings. No global temperature or continuous-input classification is implied.}
\label{tab:eos-input-scan}
\end{table}

\subsection{Audit scope and reproducibility}
The audits check the vacuum inclusion, parameter refinement, continuous temperature coverage, gap endpoint signs, curvature, free terms, and infinite tails, reconstructing the final inequalities with exact rational arithmetic where applicable. They do not independently recompute every finite quadrature contribution. The certificates use \texttt{mpmath.iv}~\cite{MpmathIntervals} and remain conditional on the reviewed integration implementation and interval engine; they are not machine-checked formal proofs. The accompanying manifest and records retain the exact domains, source hashes, precision checks, and audit tests.

\subsection{EOS bounds and the local fold mechanism}
\label{app:eos-details}
For the equilibrium EOS of Sec.~\ref{sec:eos-speed},

\begin{equation}
 c_{\mathrm{EOS}}^2\equiv\frac{dP}{d\epsilon}
 =\frac{s}{C_V}=\frac{s}{TR}.
 \label{eq:eos-speed}
\end{equation}
We retain the direct numerator and denominator bounds needed to establish a continuous interval of violation.

To establish the violation throughout a temperature interval, we take $J_s=[93.5,93.5625]\MeV$ and the same exact parameter box. We combine entropy bounds at the lower endpoint with response bounds valid throughout this interval and obtain
\begin{align}
 s(93.5\MeV)&>9.7123\times10^5\,\mathrm{MeV}^3,\\
 0<R(T)&<9.9073\times10^3\,\mathrm{MeV}^2,\\
 C_V(T)&<9.2696\times10^5\,\mathrm{MeV}^3,
 \qquad T\in J_s.
 \label{eq:eos-enclosures}
\end{align}
For each fixed set of model parameters, $R=ds/dT>0$ ensures that the entropy never falls below its lower-endpoint bound anywhere in $J_s$. Meanwhile, with $R_{\max}=9.9073\times10^3\,\mathrm{MeV}^2$ the response upper bound in Eq.~\eqref{eq:eos-enclosures}, $TR\leq(93.5625\MeV)R_{\max}$ gives a uniform upper bound on the denominator. Dividing these conservative bounds yields
\begin{equation}
 c_{\mathrm{EOS}}^2(T)>1.047>1,
 \qquad T\in J_s.
 \label{eq:eos-bound}
\end{equation}
These bounds include the true fit but do not locate the complete violation interval. The point estimate and real-time interpretation are discussed in Sec.~\ref{sec:eos-speed}.

For a numerical input-sensitivity comparison, we hold $\alpha=1$, $m_\pi=139\MeV$, and $f_\pi=92.4\MeV$ fixed and refit the vacuum parameters at each $c$ in Table~\ref{tab:eos-input-scan}. A $0.25\MeV$ temperature grid on $[80,110]\MeV$ locates candidate extrema and crossings of $s/C_V=1$, which are then refined numerically. Every sampled entropy, heat capacity, and scalar curvature is positive. Increasing the vacuum and thermal quadrature orders leaves the displayed results unchanged. This is a convergence check, not a certified enumeration of extrema or a bound over continuous input values; narrow additional components between grid points are not excluded.

The high-temperature heat-capacity fold of Sec.~\ref{sec:fit} explains how a large EOS derivative can arise while both thermal quantities stay positive. We evaluate the complete entropy over the same fitted-parameter and equilibrium-field enclosures used for the fold certificate in Appendix~\ref{app:stationary}. On its full rectangle $\mathcal C\times J$, with $J=[93.25,93.95]\MeV$, the interval evaluation, including the integration tails and restored massive free entropy, gives
\begin{equation}
 \begin{gathered}
 2.3\times10^5<\frac{s(c,T)}{\mathrm{MeV}^3}<1.6\times10^6,\\
 (c,T)\in \mathcal C\times J.
 \end{gathered}
 \label{eq:fold-entropy-positive}
\end{equation}
The exact interval $\mathcal C=[c_-,c_+]$ is specified in Appendix~\ref{app:stationary}. Thus the unique fold $(c_h,T_h)$ in this rectangle has $s_h=s(c_h,T_h)>0$, in addition to $R=R_T=0$, $a_h=R_c>0$, and $b_h=R_{TT}>0$.

The variable $c$ follows the smooth vacuum-refitted family, whereas temperature derivatives hold each fitted model fixed. With $\delta c=c-c_h$ and $\tau=T-T_h$, the ordinary fold expansion reads
\begin{equation}
 \begin{aligned}
 R(c,T)={}&a_h\delta c+\tfrac12b_h\tau^2\\
 &+O(\delta c^2+|\delta c\,\tau|+|\tau|^3).
 \end{aligned}
 \label{eq:fold-eos-expansion}
\end{equation}
At fixed temperature $T=T_h$, the response satisfies $R=a_h\delta c+O(\delta c^2)$ while the entropy satisfies $s=s_h+O(\delta c)$. We therefore obtain
\begin{equation}
 c_{\mathrm{EOS}}^2(c,T_h)
 =\frac{s_h}{T_h a_h}\frac1{\delta c}+O(1),
 \qquad\delta c\to0^+.
 \label{eq:fold-eos-enhancement}
\end{equation}
Equation~\eqref{eq:fold-entropy-positive} establishes the required positive numerator at this fold. The enhancement is local in $c$ at fixed $T_h$ and supplies no quantitative error bound at the nominal input $c=225\MeV$. The direct bounds in Eq.~\eqref{eq:eos-bound} remain the independent evidence for the nominal-fit result.

\subsection{Ideal-pion reference and its limitations}
\label{app:meson-reference}
For the scale comparison in Sec.~\ref{sec:meson-scope}, the pressure of three ideal pions and its dilute limit are

\begin{align}
 P_\pi(T)&=-3T\ip\log(1-e^{-E_\pi/T}),
 \quad E_\pi=\sqrt{p^2+m_\pi^2},\\
 P_\pi(T)&\sim3T\left(\frac{m_\pi T}{2\pi}\right)^{3/2}e^{-m_\pi/T},
 \qquad T\ll m_\pi.
 \label{eq:dilute-pion}
\end{align}
The leading derivatives are $s_\pi\sim m_\pi P_\pi/T^2$ and $C_{V,\pi}\sim m_\pi^2P_\pi/T^3$. At $T=93.5\MeV$, the full Bose integrals with $m_\pi=139\MeV$ give $s_\pi\simeq8.01\times10^5\,\mathrm{MeV}^3$ and $C_{V,\pi}\simeq2.81\times10^6\,\mathrm{MeV}^3$. Together with the nominal quark values $s\simeq9.71\times10^5\,\mathrm{MeV}^3$ and $C_V\simeq5.04\times10^5\,\mathrm{MeV}^3$, these give the illustrative additive ratio $(s+s_\pi)/(C_V+C_{V,\pi})\simeq0.53$. The dilute approximation is not used at this temperature. The pion gas has no scalar feedback or vacuum refit; this sum is not a next-order EOS prediction. Its limitations are stated in Sec.~\ref{sec:meson-scope}.

\section{Existence of the fitted thermal folds}\label{app:stationary}
\subsection{A continuous vacuum branch and a minimum over a temperature interval}
We next establish a thermal fold on the true fitted branch while propagating the computational parameter uncertainty. Let $\mathcal C=[c_-,c_+]$ contain the high-temperature heat-capacity candidate. For every $c\in\mathcal C$, the interval bounds in Appendix~\ref{app:stationary} establish a unique fitted vacuum solution in the specified local parameter range, varying smoothly with $c$.

The positive scalar root is unique and has nonzero curvature, so the equilibrium scalar field varies smoothly with the condensate input and temperature. It follows that $R(c,T)$ is continuous on $\mathcal C\times J$, where $J=[T_-,T_+]$, and differentiable in $T$. We use $D_T$ for the temperature derivative along the equilibrium scalar solution at fixed fitted parameters. To locate a stationary zero, we consider the minimum
\begin{equation}
 v(c)=\min_{T\in J}R(c,T).
\end{equation}
The minimum varies continuously with the condensate input, as follows from the bound
\begin{equation}
 |v(c)-v(c')|\leq\sup_{T\in J}|R(c,T)-R(c',T)|.
\end{equation}
Suppose $R(c_-,T_s)<0$ at one interior $T_s$, $R(c_+,T)>0$ at every $T\in J$, and $R(c,T_\pm)>0$ at both temperature endpoints for all $c\in\mathcal C$. Then the continuous minimum $v$ crosses zero at an interior $c_*$. The minimum is attained because the temperature interval is closed and bounded, and positivity at its endpoints places a minimizing $T_*$ in the interior of $J$. At this point,
\begin{equation}
 R(c_*,T_*)=0,\qquad D_TR(c_*,T_*)=0.
 \label{eq:jointstationary}
\end{equation}
Together with the three vacuum equations and the thermal gap equation, these conditions give six simultaneous equations for the fitted vacuum and thermal equilibrium solutions. The minimum argument proves that they have a simultaneous solution. Below, rigorous derivative bounds establish uniqueness, strict $D_T^2R>0$, and a nonzero derivative with respect to $c$.

\subsection{Certified ranges and numerical location}

For the certificate, we choose $c_\pm=c_0\pm2\times10^{-5}\MeV$, with the exact rational $c_0$ given in Appendix~\ref{app:stationary}, and $J=[93.25,93.95]\MeV$. The four sign bounds, valid throughout the fitted-parameter uncertainty ranges, appear in Table~\ref{tab:stationary}. The negative value occurs at $T_s\simeq93.595\MeV$, with the exact temperature specified in Appendix~\ref{app:stationary}; the argument requires only its negative sign and does not assume that this is the stationary temperature. Interval bounds covering the full temperature interval establish the bound at the upper-$c$ endpoint for every temperature in $J$.

We therefore obtain a simultaneous solution with condensate input and temperature strictly between the certified endpoints
\begin{align}
 c_-&<c_*<c_+,\\
 93.25&<T_*/\mathrm{MeV}<93.95.
 \label{eq:jointrectangle}
\end{align}
At this root $C_V=D_TC_V=0$. The rounded location in Table~\ref{tab:folds} is a numerical estimate, whereas Eq.~\eqref{eq:jointrectangle} is the certified existence region.

\begin{table}[!htbp]
\begin{ruledtabular}
\begin{tabular}{lc}
Domain&Certified bound\\
\hline
$(c_-,T_s)$&$R<-0.2265$\\
$\mathcal C\times\{T_-\}$&$R>48.03$\\
$\mathcal C\times\{T_+\}$&$R>54.12$\\
$\{c_+\}\times J$&$R>0.1149$
\end{tabular}
\end{ruledtabular}
\caption{Strict thermal sign bounds used in the stationary-zero existence proof. All bounds are for the complete equilibrium $R=C_V/T$, in $\mathrm{MeV}^2$. $\mathcal C=[c_-,c_+]$ is the certified condensate-input interval and $J=[T_-,T_+]$ the temperature interval. The bounds cover all fitted parameters in their uncertainty ranges and all temperatures in each stated domain.}
\label{tab:stationary}
\end{table}

\begin{table}[!htbp]
\begin{ruledtabular}
\begin{tabular}{lccc}
Component&$J/\mathrm{MeV}$&$k$&$a$\\
\hline
Low-$T$ $s$&$[58.1,58.175]$&3617&962\\
High-$T$ $s$&$[101,101.25]$&286909&12635\\
Low-$T$ $C_V$&$[56.51,56.56]$&430&146
\end{tabular}
\end{ruledtabular}
\caption{Additional certified ordinary folds. Each condensate interval is $\mathcal C=[c_0-2\times10^{-5},c_0+2\times10^{-5}]\MeV$, with exact $c_0$ specified in Appendix~\ref{app:uniquefold}. Bounds are strict and hold on the full rectangle. For $Y=s$, $[k]=\mathrm{MeV}^2$ and $[a]=\mathrm{MeV}$; for $Y=R$, $[k]=\mathrm{MeV}$ and $a$ is dimensionless.}
\label{tab:otherfolds}
\end{table}

\subsection{A unique nondegenerate fold}\label{sec:uniquefold}
We calculate the derivatives along the same true fitted vacuum branch. In $R_c$, physical $T$ stays fixed while all fitted parameters vary. The derivative $R_{TT}=D_T^2R$ holds each fitted model fixed and includes the scalar response. Using the construction in Appendix~\ref{app:uniquefold}, we cover the exact closed rectangle $\mathcal C\times J$ continuously and find
\begin{align}
 R_c&>11790\MeV, & R_{TT}&>515.
 \label{eq:strictfoldbounds}
\end{align}
These conservative lower bounds include changes in the fitted vacuum parameters as the condensate input varies, together with the response of the equilibrium scalar field. They apply at every condensate input and temperature in the stated rectangle.

For $c_2>c_1$, Eq.~\eqref{eq:strictfoldbounds} gives
\begin{equation}
 v(c_2)-v(c_1)\geq \kappa(c_2-c_1)>0,
\end{equation}
where $\kappa$ is the positive lower bound for $R_c$. The minimum $v(c)$ therefore has at most one zero. At fixed $c$, strict convexity in $T$ makes the stationary temperature the unique minimizer of $R(c,T)$ on $J$. Together with the interior existence result, these properties establish exactly one joint solution of $R=R_T=0$ in the fitted rectangle. Its Jacobian satisfies
\begin{equation}
 \det\frac{\partial(R,R_T)}{\partial(c,T)}
 =R_cR_{TT}>0.
 \label{eq:foldnonsingular}
\end{equation}
Since the local vacuum root is unique and the thermal gap derivative is positive, this reduced nonsingularity also establishes nonsingularity of the full six-equation system.

The endpoint signs and strict convexity give exactly two simple zeros below this local fold, one quadratic zero at it, and no zeros above it within the rectangle. The width scaling follows from Eq.~\eqref{eq:foldscaling}; no exchange of scalar minima is involved.

\subsection{Joint certification of the other three folds}\label{sec:otherfolds}

We apply the same method to the two entropy folds and the low-temperature heat-capacity fold. Each calculation gives rigorous bounds on its vacuum solution as the condensate input varies continuously, with the original pion inputs fixed at their exact values. We set $Y=s$ for an entropy fold and $Y=R$ for a heat-capacity fold. Over each rectangle $\mathcal C\times J$, we establish $Y_c\geq k>0$ and $Y_{TT}\geq a>0$ with the conservative bounds listed in Table~\ref{tab:otherfolds}. For entropy, $Y_T=R$ and $Y_{TT}=R_T$, so the joint fold equations are $s=R=0$.

The four boundary-sign conditions and positive derivatives establish one nondegenerate joint root per rectangle by the same minimum argument. The two-zero classification is local to these certified domains, not an enumeration of all thermal roots.

\subsection{Parameterized fit and response derivatives}

We center the condensate interval at the exact value $c_0=(15436081192793/68719476736)\MeV$ and set $c_\pm=c_0\pm(1/50000)\MeV$. The point at which the response is certified negative uses the exact decimal rational $T_s=93.59465552508623\MeV$. The rescaled parameter point in Sec.~\ref{sec:inputresponse} uses the exact values $(m_\pi,f_\pi,T_w)=a(139\MeV,92.4\MeV,T_s)$, where $a=225\MeV/c_-$; the shortened displayed decimals are not substituted into the certificate. Because only the third vacuum residual depends explicitly on $c$, its dependence on a reference $c_0$ is
\begin{align}
 F_{v,3}(x,c)&=(c_0/c)^3[F_{v,3}(x,c_0)+1]-1,\\
 (J_v)_{3j}(x,c)&=(c_0/c)^3(J_v)_{3j}(x,c_0).
\end{align}
The other rows are unchanged. On the original convex vacuum box, the parameterized Krawczyk inclusion has a contraction factor below $0.005592$ throughout the condensate interval and a nonsingular Jacobian. Uniqueness and the implicit-function theorem therefore give unique fitted vacuum parameters that vary continuously with the condensate input. We narrow their enclosing intervals over the full $c$ interval and at its endpoints through ten intersections with mean-value bounds, each of which retains the true root. To prove positivity at the upper-$c$ endpoint in Table~\ref{tab:stationary}, we use 47 adjoining temperature cells and mean-value bounds on $D_TR$. The temperature derivative of the free contribution is integrated directly with its infinite tail included.

The numerical simultaneous-equation solve checks convergence separately from these interval bounds.

For $x=p/T$, $a=m/T$, $w=\sqrt{x^2+a^2}$, and $n=(e^w+1)^{-1}$, the exact free slope and its temperature derivative are
\begin{align}
 R_{\mathrm{free}}&=\frac{2dT^2}{\pi^2}\int_0^\infty x^2w^2n(1-n)\,\dd x,\\
 \partial_TR_{\mathrm{free}}&=\frac{2dT}{\pi^2}\int_0^\infty x^2n(1-n)\nonumber\\
 &\quad\times[2x^2+a^2w\tanh(w/2)]\,\dd x.
\end{align}
In the equivalent energy representation $R_{\mathrm{free}}=(2dT^2/\pi^2)f(a)$,
\begin{equation}
 f(a)=\int_a^\infty z^3\sqrt{z^2-a^2}\,n_F(z)[1-n_F(z)]\,\dd z,
\end{equation}
where $n_F(z)=(e^z+1)^{-1}$. Since $f>0$ and $f'<0$ for $a>0$, the slope decreases with $m$ and increases with $T$. Its extrema over the mass and temperature box can therefore be bounded at the endpoints. We bound the temperature derivative of this slope directly over the whole mass and temperature intervals. Beyond $x=P$, the positive omitted integrands are bounded respectively by
\begin{align}
 &(x^4+a_{\max}^2x^2)e^{-x},\\
 &(2x^4+a_{\max}^2x^3+a_{\max}^3x^2)e^{-x}.
\end{align}
The integrals of these bounds are explicit exponential-polynomial expressions. On the finite domain, we use degree-seven Taylor integration with an eighth-derivative remainder. We check the formulas independently against the massless Stefan--Boltzmann limits and arbitrary-precision massive integrals. These comparisons do not supply the rigorous error bounds.

The temperature derivative along the equilibrium scalar-field solution, used in the mean-value bound at the upper condensate endpoint, follows by differentiating Eq.~\eqref{eq:dimensionlessR} with $u_T=-F_T/F_u$:
\begin{align}
 D_TR={}&s_{TT}^{\mathrm{fix}}
 +3\Lambda^4\frac{F_TF_{TT}}{F_u}
 -3\Lambda^4\frac{F_T^2F_{Tu}}{F_u^2}\nonumber\\
 &+\Lambda^4\frac{F_T^3F_{uu}}{F_u^3}.
 \label{eq:totalRT}
\end{align}
Here $s_{TT}^{\mathrm{fix}}$ is the second temperature derivative of the entropy at fixed scalar field, including the free contribution. To obtain the signs in Eq.~\eqref{eq:totalRT}, we use the mixed identity $\partial_u s_T^{\mathrm{fix}}=-\Lambda^4F_{TT}$. This term is required for the equilibrium derivative.

For the interaction integrals define dimensionless $q=r^2+v_n$, $v_n=[(2n+1)\pi T/\Lambda]^2$, $\mathcal H=g\widehat M/(q+\widehat M^2)$, and $L=\log[(q+\widehat M^2)/(q+\mu_m^2)]$. Let $\mathcal I f=\sum_{n\geq0}\int_0^\infty f\,\dd r$. Here $D=q+\widehat M^2$ and $D_0=q+\mu_m^2$ are dimensionless denominators. Partial derivatives of $\mathcal H$ and $L$ with respect to $q$ or $u$ give
\begin{align}
 F_{TT}&=-\frac{48}{\pi^2\Lambda T}\mathcal I\{v_n[-\mathcal H+(r^2-v_n)\mathcal H_q]\},\\
 F_{Tu}&=-\frac{24}{\pi^2\Lambda}\mathcal I[(r^2-v_n)\mathcal H_u],\\
 F_{uu}&=-\frac{24T}{\pi^2\Lambda}\mathcal I[r^2\mathcal H_{uu}],\\
 (s_{TT}^{\mathrm{fix}})_{\mathrm{int}}
 &=\frac{24\Lambda^3}{\pi^2T^2}\mathcal I\{v_n[-L+(r^2-5v_n)L_q\nonumber\\
 &\hspace{38mm}+2v_n(r^2-v_n)L_{qq}]\}.
\end{align}
We obtain these radial expressions by integrating by parts over the full infinite domain. To bound their tails, we use lower and upper endpoints denoted by $\min$ and $\max$, with $q\geq q_{\min}>0$, $D,D_0\geq q_{\min}$, $\widehat M\leq\mu_{m,\max}+u_{\max}$, $|g_q|\leq(1+2q)g$, and $|g_{qq}|\leq[(1+2q)^2+2]g$. These inequalities bound the absolute value of each integrand by a positive polynomial $P(q)g$, whose coefficients are stored explicitly. For every integer $k\geq0$,
\begin{equation}
 q^k e^{-(q+q^2)}\leq2^k k!\,e^{-(q+q^2)/2},
\end{equation}
so the tail bounds with half the original exponent in Appendix~\ref{app:enclosures} bound the omitted contributions to all four partial derivatives. We record the polynomial coefficients and the amplitudes of these upper bounds in exact rational arithmetic. Independent integrand differentiation and equilibrium entropy differences check the implementation separately from the interval remainder construction.

The numerical scaling comparisons shown in Fig.~\ref{fig:folds} and their precision checks are retained in the computational records.

\section{Derivatives for uniqueness and nondegeneracy}\label{app:uniquefold}
With $x_c=\dd x/\dd c$, the condensate derivative of the vacuum equations, evaluated at their true root, is
\begin{equation}
 J_v x_c=(0,0,3/c)^\mathsf{T}.
\end{equation}
We bound the derivatives of the fitted vacuum parameters by solving this linear system with the interval Jacobian and preconditioning matrix already used for the continuous fitted solution, applying a weighted Neumann series followed by intersections with mean-value bounds. Differentiating $\lambda=u_0/A_v$ then gives
\begin{equation}
 \lambda_c=\frac{(u_0)_c}{A_v}
       -\frac{u_0\,\nabla A_v\cdot x_c}{A_v^2}.
\end{equation}
To include the change of the thermal response as the vacuum parameters are refitted, we write $R=\Lambda^2\widehat R(t,\mu_m,\lambda)$, where $\widehat R$ is the dimensionless equilibrium response and $t=T/\Lambda$:
\begin{align}
 R_\Lambda&=\frac{2R-TR_T}{\Lambda},\\
 R_c&=R_{\mu_m}(\mu_m)_c+R_\lambda\lambda_c
                         +R_\Lambda\Lambda_c.
\end{align}
The partial parameter derivatives of $R$ here include the change of the equilibrium scalar field. We evaluate four additional mass-partial integrals for $F_{\mu_m}$, $F_{T\mu_m}$, $F_{u\mu_m}$, and the mass derivative of the fixed-field entropy slope. The free massive derivative is integrated separately. Holding the fitted triple fixed while differentiating would omit its required change with the condensate input.

For the second temperature derivative it is convenient to write the dimensionless complete potential as $\phi(t,u)=\Omega/\Lambda^4$, including the restored massive free term. Let $F=\phi_u$, $b=\dd u/\dd t=-F_t/F_u$, and define $D=\partial_t+b\partial_u$. This operator varies temperature and scalar field together in the equilibrium direction at each point; $b$ is held constant while taking these directional partial derivatives. The corresponding fourth-order envelope identity for the equilibrium potential is
\begin{equation}
 R_{TT}=-D^4\phi+3\frac{(D^2F)^2}{F_u}.
\end{equation}
This identity includes the second temperature derivative of the equilibrium scalar field implied by the gap equation. We obtain the two interaction integrals directly from Taylor coefficients of the determinant and gap when temperature and scalar field vary together in the specified direction, so no fourth derivative is inferred from numerical finite differences.

In both constructions, every bound covers the full parameter, temperature, and scalar-field intervals simultaneously. We integrate the analytic derivative integrands with eighth-derivative radial Taylor remainders and bound the absolute values of the omitted interaction contributions by explicitly stored positive polynomials times $e^{-q-q^2}$. The free derivatives have separate tail bounds given by polynomials times exponentials. Independent arbitrary-precision evaluations of the integrands and derivatives check the implementation, and exact-rational audits reconstruct the final enclosing intervals. These calculations use the finite-quadrature and interval-engine assumptions stated in Appendix~\ref{app:enclosures}.

For the three additional folds the exact condensate centers, in MeV, are
\begin{align}
 c_{0,s,\mathrm{low}}&=\frac{30419442432229}{137438953472},\\
 c_{0,s,\mathrm{high}}&=\frac{121642904492021}{549755813888},\\
 c_{0,R,\mathrm{low}}&=\frac{30705986496515}{137438953472}.
\end{align}
The corresponding vacuum-root contraction factors are below $0.006273$. To calculate how the entropy changes with the condensate input, we combine interval bounds on the derivatives of the fitted vacuum parameters with the parameter derivatives
\begin{align}
 s_{\mu_m}&=(s_{\mathrm{int}}^{\mathrm{fix}})_{\mu_m}
       +\Lambda\partial_m s_{\mathrm{free}}
       +\Lambda^4\frac{F_TF_{\mu_m}}{F_u},\\
 s_\lambda&=-\Lambda^4\frac{F_Tu}{\lambda^2F_u},\qquad
 s_\Lambda=\frac{3s-TR}{\Lambda},\\
 s_c&=s_{\mu_m}(\mu_m)_c+s_\lambda\lambda_c+s_\Lambda\Lambda_c.
\end{align}
On the first line, only the fixed-field term is an interaction derivative. The displayed $s_{\mu_m}$ is the total derivative after scalar equilibration and retains the explicit free derivative. We keep ten positive frequencies in the entropy integrals and additional partials, and twelve in the low-temperature heat-fold integrals. We bound the omitted infinite sums analytically, as we do the finite-domain remainders.

The sign margins near these shallow minima are small and can be obscured when the fitted parameters are bounded separately, losing their correlations. We therefore first obtain a tight interval for $Y(c_0,T)$, then integrate the certified bounds on $Y_c$ to bound its change with condensate input. This gives bounds at both condensate endpoints and at both temperature boundaries of the rectangle. At the upper-$c$ endpoint, strict convexity gives the global lower bound
\begin{equation}
 \min_{T\in J}Y(c_+,T)\geq
 Y(c_+,T_s)-\frac{|Y_T(c_+,T_s)|^2}{2a}.
 \label{eq:convexslice}
\end{equation}
Completing the square in the quadratic lower Taylor bound gives this result without requiring the stationary temperature. A stationary zero then exists because the lower-$c$ endpoint has a certified negative value at the chosen temperature, both temperature boundaries remain positive throughout the condensate interval, and the right-hand side of Eq.~\eqref{eq:convexslice} is positive. The derivative bounds over the complete rectangle prove uniqueness and give $\det\partial(Y,Y_T)/\partial(c,T)=Y_cY_{TT}>0$ at the root.

The bounds away from the central condensate input used in Appendix~\ref{sec:otherfolds} follow by integrating $k\leq Y_c\leq K$ on the complete rectangle. For $\delta_c=c_+-c_0=c_0-c_-$, rigorous lower and upper bounds at the central input give
\begin{align}
 Y(c_-,T_s)&\leq Y(c_0,T_s)_+-k\delta_c,\\
 Y(c_+,T_s)&\geq Y(c_0,T_s)_-+k\delta_c,\\
 \inf_{c\in\mathcal C}Y(c,T_\pm)&\geq Y(c_0,T_\pm)_--K\delta_c.
\end{align}
In these inequalities, subscripts $-$ and $+$ denote the lower and upper bounds of the computed interval. The notation $\inf_{c\in\mathcal C}$ denotes the greatest lower bound over all allowed condensate inputs in $\mathcal C$. We separately bound the temperature derivative at the upper condensate endpoint in Eq.~\eqref{eq:convexslice}, uniformly in the parameters. Independent audits reconstruct the Krawczyk contraction at the central input, the inequalities obtained by integrating the condensate derivative, the complete temperature-cell coverage, and the strong-convexity arithmetic. The sign bounds never use numerical convergence differences.

% Low-temperature appendix; explicit Fourier convention and distinct pole-response notation.
% Model: g(q)=exp(-q-q^2), q=Q^2/Lambda^2; mu_m is current mass/Lambda.
\section{Uniform control of the open low-temperature interval}
\label{app:phase27-low-temperature}

We now prove the low-temperature result for every point of the certified
fitted parameter box at $c=225\MeV$,
with $d=N_cN_f=6$, $\mu_m=m/\Lambda>0$, $u=\sigma/\Lambda>0$,
and $\lambda=G\Lambda^2$. All intervals for the scalar field and pole positions
below cover the entire parameter box. The symbols $\mu_{m,\pm}$ and $\Lambda_\pm$ denote the endpoints of the exact fitted intervals underlying Table~\ref{tab:box}; $u_\pm$ denote the endpoints of the field interval $U_{40}$ given below. We use the mass kernel $g(q)=e^{-q-q^2}$, with $q=Q^2/\Lambda^2$,
at vanishing chemical potential.

\subsection{A uniform thermal field enclosure}
Put $t=T/\Lambda$, $\bm r=\bm p/\Lambda$, and $r=|\bm r|$. We use the dimensionless denominator $\mathcal D(q)=q+[\mu_m+ug(q)]^2$ and define
\begin{align}
 H(q)&=\frac{g(q)[\mu_m+ug(q)]}{q+[\mu_m+ug(q)]^2},\\
 F(u,T)&=\frac{u}{\lambda}-24t\sum_{n\in\mathbb Z}
  \int\frac{d^3r}{(2\pi)^3}H\bigl(r^2+[(2n+1)\pi t]^2\bigr).
\end{align}
The vacuum residual $F(u,0)$ is obtained by replacing the Matsubara sum with a continuous Euclidean-frequency integral. With $x$ the Euclidean frequency in units of $\Lambda$, the fermionic Poisson formula and a contour shift through a dimensionless height $a=0.45$ yield
\begin{align}
 |F(u,T)-F(u,0)|&\le\frac{48I_a}{e^{a\Lambda/T}-1},
 \label{eq:phase27-gap-bound}\\
 I_a&=\int\frac{d^3r\,dx}{(2\pi)^4}
  \left|H\bigl(r^2+(x+ia)^2\bigr)\right|.
\end{align}
Here the factor 48 accounts for both signs of the nonzero Poisson harmonics.
We use a rectangular contour shift, which requires no decay on an infinite
semicircle. With
$\zeta=r^2+x^2$, the shifted kernel obeys
\begin{equation}
 |g|\le G_a(\zeta)=e^{a^2-a^4+(-1+6a^2)\zeta-\zeta^2}.
\end{equation}
We establish $\operatorname{Re}\mathcal D>0$ on the upper line for $0\le\zeta\le4$
using interval angular bounds. Beyond this range,
$\operatorname{Re}\mathcal D\ge\zeta-a^2-(\mu_{m,+}+u_+G_a)^2\ge\zeta/2$.
For $0\le y\le a$, $G_y\le G_a$.  The real axis and the vertical sides
of sufficiently wide rectangles also have positive $\operatorname{Re}\mathcal D$.
The harmonic minimum principle then excludes zeros throughout the strip,
and quartic decay makes the vertical integrals vanish. To bound $I_a$,
we use the full four-dimensional radial measure $\zeta\,d\zeta/(16\pi^2)$ with an
analytic tail. This bound also justifies exchanging the momentum integral
and harmonic sum.

For $0<T\le40\,\mathrm{MeV}$, Eq.~\eqref{eq:phase27-gap-bound} gives
$|F(T)-F(0)|<3.2495\times10^{-6}$ on
\begin{equation}
 U_{40}=[0.3904659380677783,\,0.3908659380677783],
\end{equation}
where $U_{40}$ encloses $u$. The displayed endpoints describe the interval; we use exact rational
endpoints in the certificates. At the lower endpoint, we find
$F<-7.9357\times10^{-8}$ and the upper endpoint has
$F>8.1350\times10^{-8}$, uniformly in temperature and parameters.
Since $g(\mu_m/u+g)/\mathcal D$ decreases strictly with $u>0$, $F/u$ increases
strictly.  The enclosed root is therefore unique, with $F_u>0$ there.
Together with $\Omega(u)<\Omega(-u)$ for $u>0$, this shows that it is the
global minimum within the homogeneous real-scalar channel.

\subsection{The complete pole series and its thermal derivatives}
Let $w_j$ be all zeros of $\mathcal D(w)$, counted with multiplicity, and define
$M_j=\Lambda\sqrt{-w_j}$ and $E_j(p)=\sqrt{p^2+M_j^2}$ with positive
real parts. Here $M_j$ are pole mass parameters of the analytically continued quark propagator; complex values do not describe stable particle masses. We count zeros by contour integrals uniformly over the parameter
and field boxes, then isolate them in disjoint disks using Rouch\'e's theorem.
This encloses one real pole and 26 conjugate pairs below
$\operatorname{Re}M/\Lambda=1.2$, with no additional poles in that strip.
Every enclosed pole satisfies $\operatorname{Re}M/\Lambda>0.47035$.
Consequently all roots obey
$R_j=|M_j|/\Lambda>R_0=0.47>\mu_{m,+}+u_+$.
The root equation then implies $\operatorname{Re}(w_j+w_j^2)\le0$, giving
the uniform sector estimate
\begin{equation}
 \operatorname{Re}E_j(p)\ge c_{\rm sec}\sqrt{p^2+|M_j|^2},\qquad
 c_{\rm sec}=\sin(\pi/8),
 \label{eq:phase27-sector}
\end{equation}
Here $c_{\rm sec}$ is a dimensionless constant bounding the allowed complex-energy sector, not a sound speed. We also have $|E_j|\le\sqrt{p^2+|M_j|^2}$.

The entire function $\mathcal D$ has order two and a genus-two Hadamard
product. The canonical factors include compensating exponentials that make
the infinite product converge. Taking six frequency derivatives of their
complete logarithms removes these polynomial compensators.
With $x$ a dimensionful Euclidean frequency and $p=|\mathbf p|$ fixed,
each logarithmic pole factor obeys
\begin{equation}
 \|\partial_x^6\log(x^2+E_j^2)\|_{L^1(\mathbb R)}
 \le90\pi(\operatorname{Re}E_j)^{-5}.
\end{equation}
The uniform Jensen bound $N_R(R)=O(R^4)$ counts roots up to dimensionless
mass magnitude $R$, including multiplicity. Together with the positive
lower bound on all root magnitudes, it ensures that the sum of the $L^1$
norms of these sixth frequency derivatives converges uniformly over the
parameter and field boxes at fixed $p$. Thus the absolute values can be
integrated over the real frequency axis and then summed over poles.
To express the nonzero Fourier modes, we define the dimensionful denominator
$\Delta(s)=\Lambda^2\mathcal D(s/\Lambda^2)$ and
\begin{equation}
 \ell_p(x)=\log\frac{\Delta(x^2+p^2)}{x^2+p^2+m^2},\qquad
 E_m(p)=\sqrt{p^2+m^2}.
\end{equation}
With the convention $\widehat f(k)=\int_{\mathbb R}dx\,e^{-ikx}f(x)$, we obtain
\begin{equation}
 \widehat\ell_p(k)=-\frac{2\pi}{|k|}
 \left[\sum_j e^{-|k|E_j(p)}-e^{-|k|E_m(p)}\right],
 \quad k\ne0.
 \label{eq:phase31-nonzero-fourier}
\end{equation}
Here $k$ has units of inverse energy, and the sum counts every zero with
its multiplicity. We derive Eq.~\eqref{eq:phase31-nonzero-fourier} by
transforming the six frequency derivatives at fixed $p$ and dividing by
$(ik)^6$. The Fourier transform of any polynomial ambiguity is supported
only at $k=0$. The original subtracted logarithm is Schwartz on the real
$x$ axis: it and all its derivatives decay faster than any inverse power.
We retain its zero mode as the original vacuum integral at the same field.

Only after this nonzero-mode transformation do we integrate over momentum
and sum the fermionic Poisson harmonics $k=\ell/T$, with weights
$(-1)^\ell$ for $\ell\in\mathbb Z\setminus\{0\}$. The sector bound
and quartic root count give an exponential upper bound that is integrable
over momentum and summable over poles and Poisson harmonics on each
closed, finite temperature interval bounded away from zero. These bounds justify exchanging
the sums and momentum integral for the exponential Fourier modes;
integrating the preceding coarse sixth-derivative bound over momentum
would not serve this purpose. The restored free thermal term cancels
the subtracted free branch, giving
\begin{equation}
 \Omega(T,u)-\Omega(0,u)
 =-4dT\sum_j\int_{\mathbf p}\log(1+e^{-E_j/T}).
 \label{eq:phase27-full-poles}
\end{equation}
The weights in this expression are root multiplicities, not propagator residues.
On every closed, finite temperature interval bounded away from zero,
Eq.~\eqref{eq:phase27-sector} and the uniform quartic count give exponential
upper bounds that remain integrable over momentum and summable over poles
after any fixed finite number of thermal derivatives. We may therefore
interchange root sums, momentum integrals, and fixed-field temperature
derivatives uniformly over the parameter and field boxes.
The argument takes no limit $\mu_m\to0$, $u\to0$, and does not remove
the quartic term.

\subsection{A direct response bound and the zero-temperature limit}
To bound the response directly, we denote the contribution of the $j$th
pole by $\mathcal R_j$, retaining $R_j=|M_j|/\Lambda$ for the magnitude of
its complex mass in units of the nonlocality scale. Differentiating the
entropy kernel gives
\begin{equation}
 \mathcal R_j=\frac{4d}{T^3}\int_{\mathbf p}
       \frac{E_j^2e^{-E_j/T}}{(1+e^{-E_j/T})^2}.
 \label{eq:phase27-pole-response}
\end{equation}
We bound the absolute value of the pole contribution by replacing $E_j^2$
with $|E_j|^2$ and bounding the magnitude of the denominator from below by
$(1-e^{-\operatorname{Re}E_j/T})^2$. Put
$a_j=\operatorname{Re}M_j>0$.  Equation~\eqref{eq:phase27-sector} at
$p=0$ gives $|M_j|\leq a_j/c_{\rm sec}$, and the stronger energy bounds are
\begin{equation}
 \begin{aligned}
 \bigl[\operatorname{Re}E_j(p)\bigr]^2&\geq a_j^2+c_{\rm sec}^2p^2,\\
 |E_j(p)|^2&\leq c_{\rm sec}^{-2}(a_j^2+c_{\rm sec}^2p^2).
 \end{aligned}
 \label{eq:phase36-energy-bounds}
\end{equation}
Indeed, $f_j(z)=[\operatorname{Re}\sqrt{z+M_j^2}]^2$ has
$f_j(0)=a_j^2$ and a nondecreasing derivative on $z\geq0$, starting
at $f_j'(0)=a_j^2/|M_j|^2\geq c_{\rm sec}^2$.  Integrating this derivative
gives the first bound; the triangle inequality gives the second.
The substitution $p'=c_{\rm sec}p$ contributes $c_{\rm sec}^{-3}$ from the measure
and $c_{\rm sec}^{-2}$ from the numerator.  The response envelope therefore
has prefactor $2dT^2/[\pi^2c_{\rm sec}^5(1-e^{-x})^2]$ and Bessel moment
$x^2[(x^2+12)K_2(x)+3xK_1(x)]$, with $x=a_j/T$.
This positive upper bound decreases for $x\geq7/2$. To sum the omitted
pole contributions, we integrate it against the Jensen bound on the
number of poles, using $R_j\leq a_j/(c_{\rm sec}\Lambda)$.
With the dimensionless pole cutoff $h=1.2$, omitted poles satisfy $a_j\geq h\Lambda$. The resulting tail bound uses upper incomplete Gamma functions and
contains the exponential factor $e^{-h\Lambda/T}$; to cover the
entire parameter box, we use the cutoff $h\Lambda_-/T$.
The resulting bound controls the response directly; it is not obtained
by differentiating an entropy upper bound.

The real pole's contribution to the response has the positive lower bound
\begin{align}
 L(T)&=\frac{12}{\pi^2}\sqrt{\frac\pi2}M_+^{7/2}T^{-3/2}
 \frac{e^{-M_+/T}P(T/M_+)}{(1+e^{-M_+/T})^2},\\
 P(r)&=1+4r+12r^2+12r^3,\qquad M_+=468\,\mathrm{MeV}.
\end{align}
Here $M_+=468\,\mathrm{MeV}$ is a uniform upper bound on the isolated real-pole mass over the parameter and field intervals. For each finite momentum cell, the complex-pole envelope is
$b(T)=C_{\rm cell}T^{-3}e^{-E_{\rm cell}/T}/(1-e^{-E_{\rm cell}/T})^2$,
where $C_{\rm cell}>0$ collects the cell measure and the upper bound on $|E_j|^2$, including the pole multiplicity and normalization, and $E_{\rm cell}>0$ bounds $\operatorname{Re}E_j$ from below on that cell. Both constants are independent of temperature. With $r=T/M_+$,
$rP'(r)/P(r)\le3$ follows from $3P-rP'=3+8r+12r^2>0$.
Equivalently, $T\,d\log P(T/M_+)/dT\le3$, so
\begin{equation}
 \frac{d}{dT}\log\frac{b(T)}{L(T)}
 \ge\frac1T\left(\frac{E_{\rm cell}-M_+}{T}-\frac92\right).
\end{equation}
The certificates establish $(E_{\rm cell}-M_+)/(40\,\mathrm{MeV})\geq9/2$ for every finite momentum cell. The same inequality holds for the minimum energy bound in the analytic momentum tail. This energy separation therefore extends the finite-pole comparison and its momentum tail to every $0<T\le40\,\mathrm{MeV}$.
For the omitted spectrum, positive-term bounds obtained from the half-integer recurrence of the upper incomplete Gamma function give an increasing
tail-to-$L$ ratio when $(h\Lambda_- - M_+)/T\ge3$, which holds for
$h=1.2$ throughout this interval.  Their total relative bound is
$\rho<0.900$, so
$R_{\rm fixed}(T)\ge(1-\rho)L(T)>0$.
The numerical lower bound $15.938\,\mathrm{MeV}^2$ applies only
at $T=40\,\mathrm{MeV}$.

To establish the zero-temperature limit uniformly, we construct an upper
bound on the absolute entropy without using a temperature grid. Here $s(T,u)=-\partial_T\Omega(T,u)$ is evaluated at fixed scalar field and model parameters, while $s_*(T)=s(T,u_*(T))$ is its equilibrium value, equal to the entropy density $s$ used in the main text. The supremum below covers the fitted parameter intervals and all $u\in U_{40}$.
Jensen's formula with radius ratio two bounds the number of poles by
$N_R(R)\le30R^2+8R^4$ throughout the box.  Set
$\gamma=1/4$, $\eta=\gamma\Lambda_-/T$, $B=\Lambda_+/T$, and
$J_k=\int_{R_0}^{\infty}R^ke^{-\eta R}dR$.  The sector bound, elementary
three-momentum integration, and Stieltjes integration against the pole
count then give
\begin{align}
 \sup_{\mu_m,u,\Lambda}|s(T,u)|
 &\le\frac{4dT^3\eta}{\pi^2\gamma^3(1-e^{-\eta R_0})}
 \bigl[13(30J_2+8J_4)\nonumber\\[-2pt]
 &\hspace{32mm}+B(30J_3+8J_5)\bigr].
 \label{eq:phase27-uniform-zero-entropy}
\end{align}
Here $J_k=e^{-\eta R_0}\sum_{j=0}^k[k!/(k-j)!]R_0^{k-j}\eta^{-j-1}$.
The right-hand side therefore tends to zero uniformly as $T\to0^+$.
Along the unique equilibrium solution, stationarity and $F_u>0$ give
\begin{equation}
 \frac{ds_*}{dT}=R_{\rm fixed}+\Lambda^4\frac{F_T^2}{F_u}>0,
 \qquad C_V=T\frac{ds_*}{dT}>0.
\end{equation}
Equation~\eqref{eq:phase27-uniform-zero-entropy} gives $s_*(0^+)=0$.
Strict increase then gives $s_*(T)>0$ throughout the open interval.

\clearpage
\bibliography{references}
\end{document}